\documentclass{elsart}

\usepackage{graphicx,natbib,amssymb}

\journal{New Astronomy}

\begin{document}

\begin{frontmatter}

\title{Anisotropy in the Hubble constant as observed in the \textit{HST} Extragalactic
Distance Scale Key Project results}

\author[Toronto]{M. L. McClure\corauthref{cor}\thanksref{now}}
\corauth[cor]{Corresponding author.}
\thanks[now]{Present address: Maths \& Applied Maths, University of Cape Town, 
Rondebosch 7701, South Africa.} 
\ead{mcclure@astro.utoronto.ca}
\author[Scarberia]{C. C. Dyer}
\ead{dyer@astro.utoronto.ca}
\address[Toronto]{Astronomy \& Astrophysics, University of Toronto, 50 Saint George 
St, Toronto ON 
\hspace{0mm} M5S 3H4, Canada}
\address[Scarberia]{Astronomy \& Astrophysics, University of Toronto at 
Scarborough, 1265 Military Trail, Toronto ON \hspace{0mm} M1C 1A4, Canada}

\begin{abstract}

Based on general relativity, it can be argued
that deviations from a uniform Hubble flow should be thought of as variations
in the Universe's expansion velocity field, rather than being thought of as peculiar 
velocities with respect to a uniformly expanding space.  
The aim of this paper is to use the observed motions of galaxies to map out 
variations in the Universe's expansion, and more importantly, to
investigate whether real variations in the Hubble expansion are detectable given the
observational uncertainties.
All-sky maps of the observed variation in
the expansion are produced using measurements obtained along specific lines-of-sight 
and smearing them across the sky using a Gaussian profile.  
A map is produced for the final results of
the \textit{HST} Extragalactic Distance Scale Key Project for the Hubble constant, a
comparison map is produced from a set of essentially independent data, and Monte
Carlo techniques are used to analyse the statistical significance of the variation
in the maps.
A statistically significant difference in expansion rate of 9 km
s$^{-1}$ Mpc$^{-1}$ is found to occur across the sky. 
Comparing maps of the sky at different distances appears to indicate two distinct 
sets of extrema with even stronger statistically significant variations.
Within our supercluster, variations tend to occur near the
supergalactic plane, and beyond our supercluster, variations tend to occur away from
the supergalactic plane.
Comparison with bulk flow studies shows some concordance,
yet also suggests the bulk flow studies may suffer confusion, failing to discern the
influence of multiple perturbations. 

\end{abstract}

\begin{keyword}
cosmology: cosmological parameters \sep cosmology: large-scale structure of universe 
\PACS 98.65.Dx \sep 98.80.-k

\end{keyword}

\end{frontmatter}

\section{Introduction}

Conventionally, the Hubble flow is thought of as being completely uniform and
isotropic. Deviations from a uniform Hubble flow are eliminated by imparting
objects' observed residual recessional velocities into peculiar velocities, such
that objects are thought to move with respect to a uniformly expanding space.
However, empirically it is only valid to consider the velocity field of the matter
and how everything is moving relative to everything else in the Universe. It is not
possible to infer the existence of an absolute space that expands uniformly and that
objects have peculiar velocities with respect to. Thus, deviations from a uniform
Hubble flow should properly be considered deviations in the Universe's expansion
itself. 

Interestingly, Raychaudhuri\citet{Ray55} showed that (ignoring vorticity) if a
velocity field has locally isotropic expansion, then the space is locally isotropic.
Yet we know from examples such as gravitational lensing that inhomogeneities alter
the curvature of space such that it is not locally isotropic. Thus, since space is
not locally isotropic, then the Universe's expansion can not be locally isotropic
either. Whether to conceive of the Universe expanding non-uniformly or whether to
conceive of it expanding uniformly with superimposed peculiar velocities is more than
just a conceptual issue, however. 

According to Raychaudhuri's
equation\citet{Ray55}, the existence of shear in a velocity field will lead to a 
decrease in the
volume expansion. Since inhomogeneities should introduce tidal forces and shear the
velocity field, then the existence of overdensities and underdensities in the
Universe should lead to shear throughout the Universe that decreases the Universe's
volume expansion compared with that of a homogeneous universe.
This effect should only be significant when measured locally in the vicinity of an 
inhomogeneity: the global influence should be quite small.
Raychaudhuri's equation also shows that the existence of vorticity (and also velocity 
dispersion in the Newtonian version) will lead to an increase in the volume expansion.
When structures start to collapse in the Universe and eventually become supported by 
vorticity or velocity dispersion, those regions of space cease shrinking, which can 
lead to an increase in the global expansion of the Universe.
Thus, it is important to consider the influence inhomogeneities may have on the 
Universe's expansion.

The Cosmological Principle---that the Universe is homogeneous and isotropic---is
generally assumed to hold, since averaged over large enough scales the Universe will
appear homogeneous. However, general relativity is needed to understand not only
small dense systems, but large diffuse systems such as the Universe, and according
to Einstein's field equations, the spacetime corresponding to a homogeneous universe
can not be used to represent a spatially-averaged inhomogeneous universe. This is
because Einstein's field equations do not equate the spacetime to the mass-energy
distribution directly. The energy-momentum tensor $T_{ab}$ depends on the Ricci
tensor $R_{ab}$ and scalar $R$, which stem from taking derivatives of the metric
tensor $g_{ab}$, with Einstein's equations equating 
\[ R_{ab} - \frac{1}{2} R g_{ab}= \kappa T_{ab} . \] 
If the left-hand side of the field equations for a homogeneous
universe is equated to the spatially-averaged mass-energy of an inhomogeneous
universe, there will generally be a discrepancy between the two sides of the field
equations, which will act like a cosmological constant and either accelerate or
decelerate the universe's expansion from that expected for a homogeneous universe.
Thus, even if the Universe may look homogeneous on large enough scales, assuming the
Universe to expand uniformly is ultimately misleading. Several 
researchers have suggested this effect may even explain the Universe's 
apparent acceleration 
\citep[reported by Perlmutter et al.][]{Per99} as being due to structure 
formation---Bildhauer and Futamase\citet{Bil91}, Bene et al.\citet{Ben03}, and Kolb et 
al.\citet{Kolb05}---although Russ et al.\citet{Russ97} argue that the effect of
inhomogeneities should be small. 

Also, conceiving of the Universe's expansion as uniform and assigning the galaxies
peculiar velocities, bulk flow studies such as that of Hudson et al.\citet{Hud04}
have continued to find that the peculiar velocities with respect to the Cosmic
Microwave Background (CMB) frame are correlated such that volumes of space of order
100 Mpc in radius are moving with bulk velocities of approximately 300--700 km
s$^{-1}$.  This suggests inhomogeneities significantly perturb the velocity field of
the Universe.   The existence of the Universe's large-scale structure of voids
and superclusters suggests the voids are underdense regions that have been
decelerated less due to gravity so they have ballooned up into roughly spherical
regions without undergoing structure formation, while the superclusters are
overdense regions where gravity has overcome the Universe's expansion
such that they have reached turnaround and collapsed in their densest regions. 

Moffat and Tatarski\citet{Mof95} looked at what observational effects we would
theoretically observe if we were to inhabit a local void. Via comparison of their
theoretical curves with a survey of redshift-distance determinations, they found the
data were better fit by a model with a local void than by a homogeneous universe.
Zehavi et al.\citet{Zeh98} used 44 type Ia supernova $H_0$ values to show that we
may just inhabit an underdense region of the Universe (where the expansion in the
velocity field has been slowed less due to gravity than in more dense regions of the
Universe). Referring to fig.\ 4 of Freedman et al.\citet{Fre01}, it appears that
the $H_0$ values tend to fall off beyond a distance of 100 Mpc, which suggests the
Universe may be expanding faster locally. A here-there difference in the Universe's
expansion could be an alternative to the notion of a now-then difference, which is
the assumption the Universe's supposed acceleration \citep[Perlmutter et 
al.][]{Per99} rests on, so it is important to account for the possible influence of
inhomogeneities on the Universe's expansion if the cosmological parameters are to be
properly determined. 

Thus, in this paper we will not assume the existence of a uniform spatial expansion
with peculiar velocities superimposed. We will use $H_0$ values measured along
different lines-of-sight to see whether local variation in $H_0$ exists, and to
produce all-sky maps of the observed variation across the sky. If more variation
exists in the maps than should be expected due to measurement errors in the data,
and if the high and low values of $H_0$ are correlated in position on the sky, then
this will be taken as evidence that the expansion is indeed locally anisotropic
across the sky. 
Since bulk flow studies find bulk flows of a few hundred km s$^{-1}$ on 100 Mpc
scales, which is predicted depending on the cosmological model \citep[e.g. 
see Zaroubi][]{Zar02}, and bulk flows only show the net flow of a sample volume rather 
than the individual variations in the velocity field, then it would be expected that 
variations in $H_0$ observed on this scale should be at least a few km 
s$^{-1}$ Mpc$^{-1}$.
                                                                                                                                                                                                                  
While it is easy enough to measure how fast objects are expanding away from us via
redshifts, it is the determination of accurate distances that is problematic in the
determination of $H_0$. Historically, the errors in $H_0$ have been so great that it
would be difficult to study real variation in the Universe's expansion rate. The
most accurate work to date to study $H_0$ is the \textit{HST} Extragalactic Distance
Scale Key Project \citep[Freedman et al.][hereafter the \textit{HST} Key
Project]{Fre01}, which yielded distances accurate enough for a meaningful study of real
variation in $H_0$, especially since most of the errors are systematic and shared by
all the $H_0$ determinations. Thus, we will map the directional variation in $H_0$
using the \textit{HST} Key Project data, comparing with a second set of data to
examine whether the same general trend is observed. The data selection will be
discussed in Sect.\ 2. The technique used to generate all-sky $H_0$ maps and study
the significance of the variations and the impact of distance will be outlined in
Sect.\ 3. In Sect.\ 4 the results will be examined from various frames of reference
and a comparison with bulk flow studies will be made. 

\section{Selection of Data}
                                                                                                                                                                                                                  
\subsection{\textit{HST} Key Project data}
                                                                                                                                                                                                                  
\begin{table}
\caption{\textit{HST} Key Project $H_0$ Data}
\begin{tabular}{ccccc}
\hline
Object ID & $\alpha$ & $\delta$ & $H_0$ & \\
or Cluster & (hours) & (degrees) & (km/s/Mpc) & Reference \\
\hline
SN 1990O & 17.15 & $+$16.2 & 67.3 $\pm$ 2.3 & Freed.\ \cite{Fre01} \\
SN 1990T & 19.59 & $-$56.2 & 75.6 $\pm$ 3.1 & Freed.\ \cite{Fre01} \\
SN 1990af & 21.35 & $-$62.4 & 75.8 $\pm$ 2.8 & Freed.\ \cite{Fre01} \\
SN 1991S & 10.29 & $+$22.0 & 69.8 $\pm$ 2.8 & Freed.\ \cite{Fre01} \\
SN 1991U & 13.23 & $-$26.1 & 83.7 $\pm$ 3.4 & Freed.\ \cite{Fre01} \\
SN 1991ag & 20.00 & $-$55.2 & 73.7 $\pm$ 2.9 & Freed.\ \cite{Fre01} \\
SN 1992J & 10.09 & $-$26.4 & 74.5 $\pm$ 3.1 & Freed.\ \cite{Fre01} \\
SN 1992P & 12.42 & $+$10.2 & 64.8 $\pm$ 2.2 & Freed.\ \cite{Fre01} \\
SN 1992ae & 21.28 & $-$61.3 & 81.6 $\pm$ 3.4 & Freed.\ \cite{Fre01} \\
SN 1992ag & 13.24 & $-$23.5 & 76.1 $\pm$ 2.7 & Freed.\ \cite{Fre01} \\
SN 1992al & 20.46 & $-$51.2 & 72.8 $\pm$ 2.4 & Freed.\ \cite{Fre01} \\
SN 1992aq & 23.04 & $-$37.2 & 64.7 $\pm$ 2.4 & Freed.\ \cite{Fre01} \\
SN 1992au & 00.10 & $-$49.6 & 69.4 $\pm$ 2.9 & Freed.\ \cite{Fre01} \\
SN 1992bc & 03.05 & $-$39.3 & 67.0 $\pm$ 2.1 & Freed.\ \cite{Fre01} \\
SN 1992bg & 07.42 & $-$62.3 & 70.6 $\pm$ 2.4 & Freed.\ \cite{Fre01} \\
SN 1992bh & 04.59 & $-$58.5 & 66.7 $\pm$ 2.3 & Freed.\ \cite{Fre01} \\
SN 1992bk & 03.43 & $-$53.4 & 73.6 $\pm$ 2.6 & Freed.\ \cite{Fre01} \\
SN 1992bl & 23.15 & $-$44.4 & 72.7 $\pm$ 2.6 & Freed.\ \cite{Fre01} \\
SN 1992bo & 01.22 & $-$34.1 & 69.7 $\pm$ 2.4 & Freed.\ \cite{Fre01} \\
SN 1992bp & 03.36 & $-$18.2 & 76.3 $\pm$ 2.6 & Freed.\ \cite{Fre01} \\
SN 1992br & 01.45 & $-$56.1 & 67.2 $\pm$ 3.1 & Freed.\ \cite{Fre01} \\
SN 1992bs & 03.29 & $-$37.2 & 67.8 $\pm$ 2.8 & Freed.\ \cite{Fre01} \\
SN 1993B & 10.35 & $-$34.3 & 69.8 $\pm$ 2.4 & Freed.\ \cite{Fre01} \\
SN 1993O & 13.31 & $-$33.1 & 65.9 $\pm$ 2.1 & Freed.\ \cite{Fre01} \\
SN 1993ag & 10.03 & $-$35.3 & 69.6 $\pm$ 2.4 & Freed.\ \cite{Fre01} \\
SN 1993ah & 23.52 & $-$27.6 & 71.9 $\pm$ 2.9 & Freed.\ \cite{Fre01} \\
SN 1993ac & 05.46 & $+$63.2 & 72.9 $\pm$ 2.7 & Freed.\ \cite{Fre01} \\
SN 1993ae & 01.29 & $-$01.6 & 75.6 $\pm$ 3.1 & Freed.\ \cite{Fre01} \\
SN 1994M & 12.31 & $+$00.4 & 74.9 $\pm$ 2.6 & Freed.\ \cite{Fre01} \\
\hline
\end{tabular}
\end{table}

\setcounter{table}{0}
\begin{table}
\caption{-- \textit{continued}}
\begin{tabular}{ccccc}
\hline
Object ID & $\alpha$ & $\delta$ & $H_0$ & \\
or Cluster & (hours) & (degrees) & (km/s/Mpc) & Reference \\
\hline
SN 1994Q & 16.50 & $+$40.3 & 68.0 $\pm$ 2.7 & Freed.\ \cite{Fre01} \\
SN 1994S & 12.31 & $+$29.1 & 72.5 $\pm$ 2.5 & Freed.\ \cite{Fre01} \\
SN 1994T & 13.21 & $-$02.1 & 71.5 $\pm$ 2.6 & Freed.\ \cite{Fre01} \\
SN 1995ac & 22.45 & $-$08.5 & 78.8 $\pm$ 2.7 & Freed.\ \cite{Fre01} \\
SN 1995ak & 02.45 & $+$03.1 & 80.9 $\pm$ 2.8 & Freed.\ \cite{Fre01} \\
SN 1996C & 13.50 & $+$49.2 & 66.3 $\pm$ 2.5 & Freed.\ \cite{Fre01} \\
SN 1996bl & 00.36 & $+$11.2 & 78.7 $\pm$ 2.7 & Freed.\ \cite{Fre01} \\
Abell 1367 & 11.74 & $+$19.8 & 75.2 $\pm$ 12.5 & Freed.\ \cite{Fre01} \\
Abell 2197 & 16.47 & $+$40.9 & 77.2 $\pm$ 12.5 & Sak.\ \cite{Sak01} \\
Abell 262 & 01.88 & $+$36.1 & 70.9 $\pm$ 11.8 & Freed.\ \cite{Fre01} \\
Abell 2634 & 23.64 & $+$27.0 & 77.7 $\pm$ 12.4 & Freed.\ \cite{Fre01} \\
Abell 3574 & 13.82 & $-$30.3 & 76.2 $\pm$ 12.2 & Freed.\ \cite{Fre01} \\
Abell 400 & 02.96 & $+$06.6 & 79.3 $\pm$ 12.6 & Freed.\ \cite{Fre01} \\
Antlia  & 10.50 & $-$35.3 & 68.8 $\pm$ 11.3 & Freed.\ \cite{Fre01} \\
Cancer  & 08.35 & $+$21.0 & 67.1 $\pm$ 11.0 & Freed.\ \cite{Fre01} \\
Cen 30 & 12.77 & $-$41.0 & 75.8 $\pm$ 12.8 & Freed.\ \cite{Fre01} \\
Cen 45 & 12.80 & $-$40.4 & 70.7 $\pm$ 11.9 & Freed.\ \cite{Fre01} \\
Coma  & 13.00 & $+$28.0 & 83.5 $\pm$ 13.4 & Freed.\ \cite{Fre01} \\
Eridanus  & 00.50 & $-$21.5 & 77.6 $\pm$ 12.9 & Freed.\ \cite{Fre01} \\
ESO 508 & 13.17 & $-$23.1 & 79.8 $\pm$ 13.0 & Freed.\ \cite{Fre01} \\
Fornax  & 03.64 & $-$35.5 & 92.2 $\pm$ 15.3 & Freed.\ \cite{Fre01} \\
Hydra  & 10.61 & $-$27.5 & 69.6 $\pm$ 11.1 & Freed.\ \cite{Fre01} \\
MDL 59 & 22.01 & $-$32.2 & 73.6 $\pm$ 11.8 & Freed.\ \cite{Fre01} \\
NGC 3557 & 11.17 & $-$37.5 & 85.0 $\pm$ 14.4 & Freed.\ \cite{Fre01} \\
NGC 383 & 01.12 & $+$32.4 & 73.9 $\pm$ 11.9 & Freed.\ \cite{Fre01} \\
NGC 507 & 01.39 & $+$33.3 & 84.9 $\pm$ 13.5 & Freed.\ \cite{Fre01} \\
Pavo  & 21.23 & $-$57.8 & 124.4 $\pm$ 19.0 & Freed.\ \cite{Fre01} \\
Pavo 2 & 21.23 & $-$57.8 & 86.3 $\pm$ 14.2 &  Sak.\ \cite{Sak01} \\
Pegasus  & 23.34 & $+$08.2 & 66.4 $\pm$ 10.7 & Freed.\ \cite{Fre01} \\
\hline
\end{tabular}
\end{table}

\setcounter{table}{0}
\begin{table}
\caption{-- \textit{continued}}
\begin{tabular}{ccccc}
\hline
Object ID & $\alpha$ & $\delta$ & $H_0$ & \\
or Cluster & (hours) & (degrees) & (km/s/Mpc) & Reference \\
\hline
Ursa Major & 11.95 & $+$49.3 & 54.8 $\pm$ 8.6 & Freed.\ \cite{Fre01} \\
Dorado  & 04.27 & $-$55.6 & 81.9 $\pm$ 8.7 & Freed.\ \cite{Fre01} \\
Hydra I & 10.61 & $-$27.5 & 82.8 $\pm$ 8.4 & Freed.\ \cite{Fre01} \\
Abell S753 & 14.06 & $-$34.0 & 87.5 $\pm$ 7.9 & Freed.\ \cite{Fre01} \\
GRM 15 & 20.05 & $-$55.5 & 95.6 $\pm$ 10.0 & Freed.\ \cite{Fre01} \\
Abell 3574 & 13.82 & $-$30.3 & 92.0 $\pm$ 10.0 & Freed.\ \cite{Fre01} \\
Abell 194 & 01.53 & $-$01.5 & 91.3 $\pm$ 7.5 & Freed.\ \cite{Fre01} \\
Abell S639 & 10.68 & $-$46.2 & 109.7 $\pm$ 9.9 & Freed.\ \cite{Fre01} \\
Coma  & 13.00 & $+$28.0 & 83.2 $\pm$ 6.0 & Freed.\ \cite{Fre01} \\
DC 2345--28 & 23.75 & $-$28.0 & 83.2 $\pm$ 6.4 & Freed.\ \cite{Fre01} \\
Abell 539 & 05.28 & $+$06.5 & 86.2 $\pm$ 6.5 & Freed.\ \cite{Fre01} \\
Abell 3381 & 06.17 & $-$33.6 & 88.9 $\pm$ 8.3 & Freed.\ \cite{Fre01} \\
NGC 4373 & 12.42 & $-$39.8 & 99.9 $\pm$ 11.2 & Freed.\ \cite{Fre01} \\
Abell 262 & 01.88 & $+$36.1 & 69.0 $\pm$ 7.7 & Freed.\ \cite{Fre01} \\
Abell 3560 & 13.53 & $-$33.2 & 78.1 $\pm$ 8.7 & Freed.\ \cite{Fre01} \\
Abell 3565 & 13.61 & $-$34.0 & 70.2 $\pm$ 7.8 & Freed.\ \cite{Fre01} \\
Abell 3742 & 21.11 & $-$47.1 & 70.0 $\pm$ 7.8 & Freed.\ \cite{Fre01} \\
Coma  & 13.00 & $+$28.0 & 70.3 $\pm$ 17.9 & Freed.\ \cite{Fre01} \\
\hline
\end{tabular}
\end{table}

The \textit{HST} Key Project data for $H_0$ have been selected as the primary data
set. This set offers a reasonably large set of values that is distributed nicely for
the purpose of making all-sky maps. Also, despite the fact that the $H_0$ values
depend on the data of other researchers, the values were all analysed by the
\textit{HST} Key Project team to be consistent with each other. Thus, the systematic
errors should be similar in most cases to minimize the effect on the study of
variations in $H_0$ so that most of the uncertainty in the relative values of $H_0$
will just be in the random errors. 
                                                                                                                                                                                                                  
There are 74 values that were published in the final \textit{HST} Key Project paper
\citep[Freedman et al.][]{Fre01} as well as 2 values that were obtained from 
Sakai
\citet{Sak01}. The data stem from the use of 4 different methods to obtain $H_0$:
the Tully-Fisher relation, the fundamental plane relation, surface brightness
fluctuations, and type Ia supernovae. 
                                                                                                                                                                                                                  
In all cases, the $H_0$ values used are those for which the recessional velocities
have been corrected to the CMB frame. That way the $H_0$ values are all in one frame
of reference, rather than making corrections for infalls, which would be contrary to
our purposes. Thus, the most recent distances from Freedman et al.\ and Sakai were
used, and where $H_0$ values corresponding to non-CMB recessional velocities were
reported, the CMB-frame recessional velocities from the former papers of Ferrarese
et al.\citet{Fer00} and Sakai et al.\citet{Sak00} were used to yield the
CMB-frame $H_0$ values. 
                                                                                                                                                                                                                  
The 2 values obtained from Sakai\citet{Sak01} had originally appeared in Sakai et
al.\citet{Sak00} but did not appear with modified distances in Freedman et 
al.\citet{Fre01}. One of these values was for Pavo 2, which is a component that was
separated from the Pavo Cluster due to some of the galaxies yielding different
Tully-Fisher distances, but since the galaxies had similar recessional velocities,
they yielded quite different $H_0$ values.  For our purposes, it only seems fair to
include the values for both Pavo components. 
                                                                                                                                                                                                                  
Reported in Table 1 are the objects used, their celestial co-ordinates, and their
$H_0$ values with corresponding 1--$\sigma$ random errors. The celestial
co-ordinates for the objects were obtained via \textsc{SIMBAD}.

\subsection{Comparison data}
                                                                                                                                                                                                                  
\begin{table}
\caption{Comparison $H_0$ Data}
\begin{tabular}{ccccc}
\hline
Object ID & $\alpha$ & $\delta$ & $H_0$ & \\
or Cluster & (hours) & (degrees) & (km/s/Mpc) & Reference \\
\hline
Coma &  13.00 & $+$28.0 & 71 $\pm$ 30 & Her.\ \cite{Her95} \\
Coma &  13.00 & $+$28.0 & 78 $\pm$ 11 & Whit.\ \cite{Whi95} \\
Virgo &  12.50 & $+$13.2 & 80 $\pm$ 16 & Zas.\ \cite{Zas96} \\
NGC 7331 &  22.62 & $+$34.4 & 70 $\pm$ 14 & Zas.\ \cite{Zas96} \\
Virgo &  12.50 & $+$13.2 & 87 $\pm$ 7 & Ford \cite{Ford96} \\
Fornax &  03.64 & $-$35.5 & 73 $\pm$ 5 & Ford \cite{Ford96} \\
NGC 5846 &  15.11 & $+$01.6 & 65 $\pm$ 8 & Forb.\ \cite{For96} \\
NGC 1365 &  03.56 & $-$36.1 & 75 $\pm$ 5 & Mad.\ \cite{Mad96} \\
Coma &  13.00 & $+$28.0 & 75 $\pm$ 6 & Gregg \cite{Gre97} \\
IC 4051 &  13.00 & $+$28.0 & 68 $\pm$ 6 & Baum \cite{Bau97} \\
Coma &  13.00 & $+$28.0 & 70 $\pm$ 7 & Hjor.\ \cite{Hjo97} \\
NGC 4889 &  13.00 & $+$28.0 & 85 $\pm$ 10 & Jen.\ \cite{Jen97} \\
NGC 3309 &  10.61 & $-$27.5 & 46 $\pm$ 5 & Jen.\ \cite{Jen97} \\
NGC 4881 &  13.00 & $+$28.0 & 71 $\pm$ 11 & Thom.\ \cite{Tho97} \\
Abell 2256 &  17.06 & $+$78.7 & 72 $\pm$ 22 & Myers \cite{Mye97} \\
Coma &  13.00 & $+$28.0 & 67 $\pm$ 26 & Myers \cite{Mye97} \\
Abell 262 &  01.88 & $+$36.1 & 82 $\pm$ 8 & Lauer \cite{Lau98} \\
Abell 3560 &  13.53 & $-$33.2 & 86 $\pm$ 7 & Lauer \cite{Lau98} \\
Abell 3565 &  13.61 & $-$34.0 & 83 $\pm$ 6 & Lauer \cite{Lau98} \\
Abell 3742 &  21.11 & $-$47.1 & 78 $\pm$ 6 & Lauer \cite{Lau98} \\
Coma &  13.00 & $+$28.0 & 60 $\pm$ 11 & Sal.\ \cite{Sal98} \\
PGC 14638 &  04.20 & $-$32.9 & 60 $\pm$ 15 & Pat.\ \cite{Pat98} \\
PGC 39724 &  12.33 & $+$29.6 & 44 $\pm$ 15 & Pat.\ \cite{Pat98} \\
PGC 51233 &  14.34 & $+$03.9 & 60 $\pm$ 15 & Pat.\ \cite{Pat98} \\
PGC 00218 &  00.05 & $+$16.1 & 58 $\pm$ 15 & Pat.\ \cite{Pat98} \\
PGC 10208 &  02.70 & $+$00.4 & 60 $\pm$ 15 & Pat.\ \cite{Pat98} \\
PGC 35164 &  11.44 & $+$43.6 & 59 $\pm$ 15 & Pat.\ \cite{Pat98} \\
PGC 43798 &  12.89 & $+$02.2 & 39 $\pm$ 15 & Pat.\ \cite{Pat98} \\
Fornax &  03.64 & $-$35.5 & 74 $\pm$ 5 & Tul.\ \cite{Tul00} \\
\hline
\end{tabular}
\end{table}

\setcounter{table}{1}
\begin{table}
\caption{-- \textit{continued}}
\begin{tabular}{ccccc}
\hline
Object ID & $\alpha$ & $\delta$ & $H_0$ & \\
or Cluster & (hours) & (degrees) & (km/s/Mpc) & Reference \\
\hline
Ursa Major &  11.50 & $+$55.0 & 59 $\pm$ 6 & Tul.\ \cite{Tul00} \\
Pisces Fil.\ &  01.12 & $+$32.4 & 79 $\pm$ 2 & Tul.\ \cite{Tul00} \\
Coma &  13.00 & $+$28.0 & 83 $\pm$ 2 & Tul.\ \cite{Tul00} \\
Abell 1367 &  11.74 & $+$19.8 & 77 $\pm$ 2 & Tul.\ \cite{Tul00} \\
Antlia &  10.50 & $-$35.3 & 86 $\pm$ 3 & Tul.\ \cite{Tul00} \\
Cen 30 &  12.77 & $-$41.0 & 83 $\pm$ 3 & Tul.\ \cite{Tul00} \\
Pegasus &  23.34 & $+$08.2 & 77 $\pm$ 3 & Tul.\ \cite{Tul00} \\
Hydra I &  10.61 & $-$27.5 & 70 $\pm$ 2 & Tul.\ \cite{Tul00} \\
Cancer &  08.35 & $+$21.0 & 80 $\pm$ 2 & Tul.\ \cite{Tul00} \\
Abell 400 &  02.96 & $+$06.6 & 76 $\pm$ 2 & Tul.\ \cite{Tul00} \\
Abell 2634 &  23.64 & $+$27.0 & 70 $\pm$ 2 & Tul.\ \cite{Tul00} \\
Abell 262 &  01.88 & $+$36.1 & 77 $\pm$ 4 & Jen.\ \cite{Jen01} \\
Abell 496 &  04.56 & $-$13.2 & 74 $\pm$ 2 & Jen.\ \cite{Jen01} \\
Abell 779 &  09.33 & $+$33.8 & 73 $\pm$ 2 & Jen.\ \cite{Jen01} \\
Abell 1060 &  10.61 & $-$27.5 & 74 $\pm$ 4 & Jen.\ \cite{Jen01} \\
Abell 1656(a) &  12.99 & $+$28.0 & 79 $\pm$ 3 & Jen.\ \cite{Jen01} \\
Abell 1656(b) &  12.99 & $+$28.0 & 82 $\pm$ 3 & Jen.\ \cite{Jen01} \\
Abell 2199 &  16.48 & $+$39.6 & 71 $\pm$ 2 & Jen.\ \cite{Jen01} \\
Abell 2666 &  23.85 & $+$27.1 & 68 $\pm$ 2 & Jen.\ \cite{Jen01} \\
Abell 3389 &  06.36 & $-$65.0 & 70 $\pm$ 2 & Jen.\ \cite{Jen01} \\
Abell 3565 &  13.61 & $-$34.0 & 78 $\pm$ 4 & Jen.\ \cite{Jen01} \\
Abell 3581 &  14.12 & $-$27.2 & 74 $\pm$ 2 & Jen.\ \cite{Jen01} \\
Abell 3656 &  19.98 & $-$38.3 & 72 $\pm$ 3 & Jen.\ \cite{Jen01} \\
Abell 3742 &  21.11 & $-$47.1 & 81 $\pm$ 4 & Jen.\ \cite{Jen01} \\
NGC 4073 &  12.07 & $+$01.9 & 64 $\pm$ 2 & Jen.\ \cite{Jen01} \\
NGC 4709 &  12.83 & $-$41.4 & 107 $\pm$ 7 & Jen.\ \cite{Jen01} \\
NGC 5193 &  13.53 & $-$33.2 & 85 $\pm$ 6 & Jen.\ \cite{Jen01} \\
Coma &  13.00 & $+$28.0 & 71 $\pm$ 8 & Liu \cite{Liu01} \\
\hline
\end{tabular}
\end{table}

In an effort to test whether variations detected in the \textit{HST} Key Project
data exist independently of this data set, another set of data has been compiled.
This data set has been constructed by conducting a literature search for
determinations of $H_0$ and using papers that report measurements of $H_0$ along
individual lines-of-sight, include uncertainties, and for which the observed objects
are within the distance range of the \textit{HST} Key Project data. This yields 57
values, which will hereafter be referred to as the comparison values. Reported in
Table 2 are the objects used, celestial co-ordinates (obtained via \textsc{SIMBAD}),
and $H_0$ values with their reported errors. 
                                                                                                                                                                                                                  
It should be noted that the comparison values stem from a mixed bag of methods and
different researcher analysis; thus, the potential for this set to be corrupted by
systematics between values is greater than in the \textit{HST} Key Project set.
Also, various infall corrections have been made, and these are not always clearly
stated in the papers, so these values may not consistently be in the CMB frame. One
of the comparison values actually stems from a \textit{HST} Key Project Cepheid
distance value, while 4 of the \textit{HST} Key Project values depend on comparison
data surface brightness fluctuation distances, but the data sets are essentially
independent.

\section{Contour mapping}
                                                                                                                                                                                                                  
\subsection{Technique}
                                                                                                                                                                                                                  
A method is required to generate an all-sky map based on a set of values located at
specific positions on the sky. One method would be to fit spherical harmonics;
however, this would require more higher order terms than could ever be convenient in
order to not force structure into the map from the lower order terms. Our chosen
method is to smear the $H_0$ values over the sky using a Gaussian profile for each
data point. 
                                                                                                                                                                                                                  
The Gaussian smearing method involves laying out a grid on the sky and calculating
weighted mean values of $H_0$ at each grid point, weighting each actual data point
in the average according to its angular separation $\theta$ from the grid point such 
that the weightings fall off as a Gaussian.
Thus, the weighting $W$ of each data point is given by
\[
W = \frac{1}{\sqrt{2\pi}\sigma}e^{-\theta^2/(2\sigma^2)}
\]
with the standard deviation $\sigma$ controlling how 
broad the smearing is.
Contours of constant $H_0$ are then interpolated within the grid of averaged $H_0$
values to generate contour maps of $H_0$. The values are weighted only by their
separations, not their uncertainties (the type Ia supernova values are more distant
and have smaller uncertainties, but the effect of distance on the map will be
specifically explored in Sect.\ 3.3). 
                                                                                                                                                                                                                  
Gaussian smearing succeeds in creating an all-sky map from a sample of data points
associated with specific positions on the sky, and it also averages out the impact
of errors associated with individual data points so that variations correlated with
directions on the sky have the opportunity to manifest themselves. This will be
sufficient for studying large-scale variations in $H_0$, although there is no hope
of studying any variations that do not have a large angular extent, as there is
insufficient sky coverage. It should be kept in mind that this method will also
smear out the extrema for any actual variation though, so the range of any real
variation in the maps will be diminished somewhat. 
                                                                                                                                                                                                                  
For the contour maps presented in this paper, while the angular separations are
calculated in spherical co-ordinates (using the dot product of unit vectors for the
angular positions of the grid points and data points), the grid points are
positioned for a cylindrical projection of the sky and are then mapped out in the
form of a sinusoidal projection. The sinusoidal projection is a pseudo-cylindrical
projection that preserves areas by keeping latitude lines parallel but shortening
their length longitudinally according to the sine of the polar angle (or cosine of
the declination). 
                                                                                                                                                                                                                  
Unless stated otherwise, the grid points are set $1^\circ$ apart in right ascension
and $1^\circ$ apart in declination, as this appears to be a fine enough grid for the
purpose of interpolating contours, and the Gaussian weighting profile falls off with
a standard deviation of $25^\circ$, as this is approximately the typical separation
of the real data points and is a sufficiently broad smearing to fill in the holes in
the distribution of data on the sky. 

\begin{figure}
\includegraphics[width=\hsize]{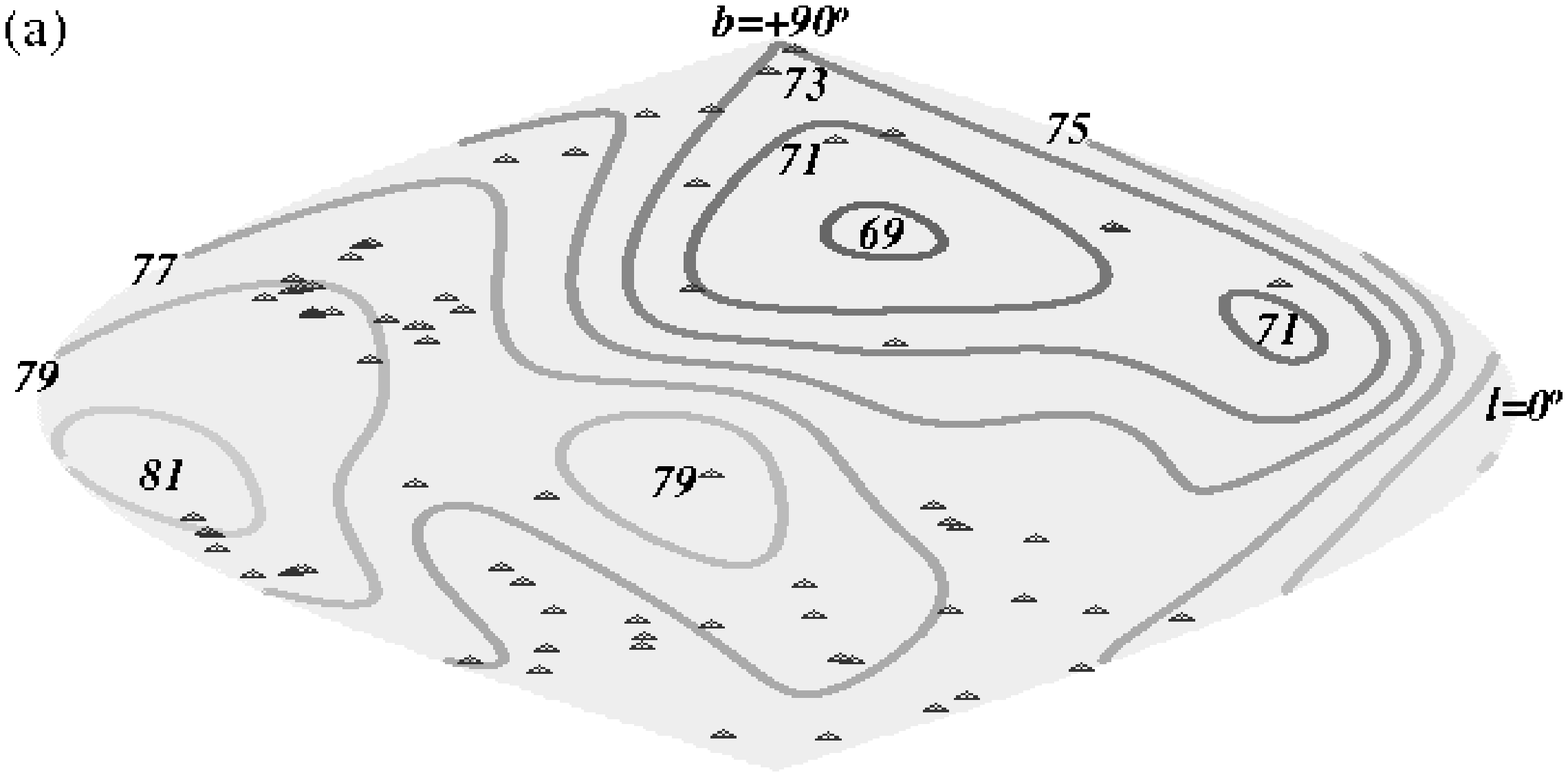}
\includegraphics[width=\hsize]{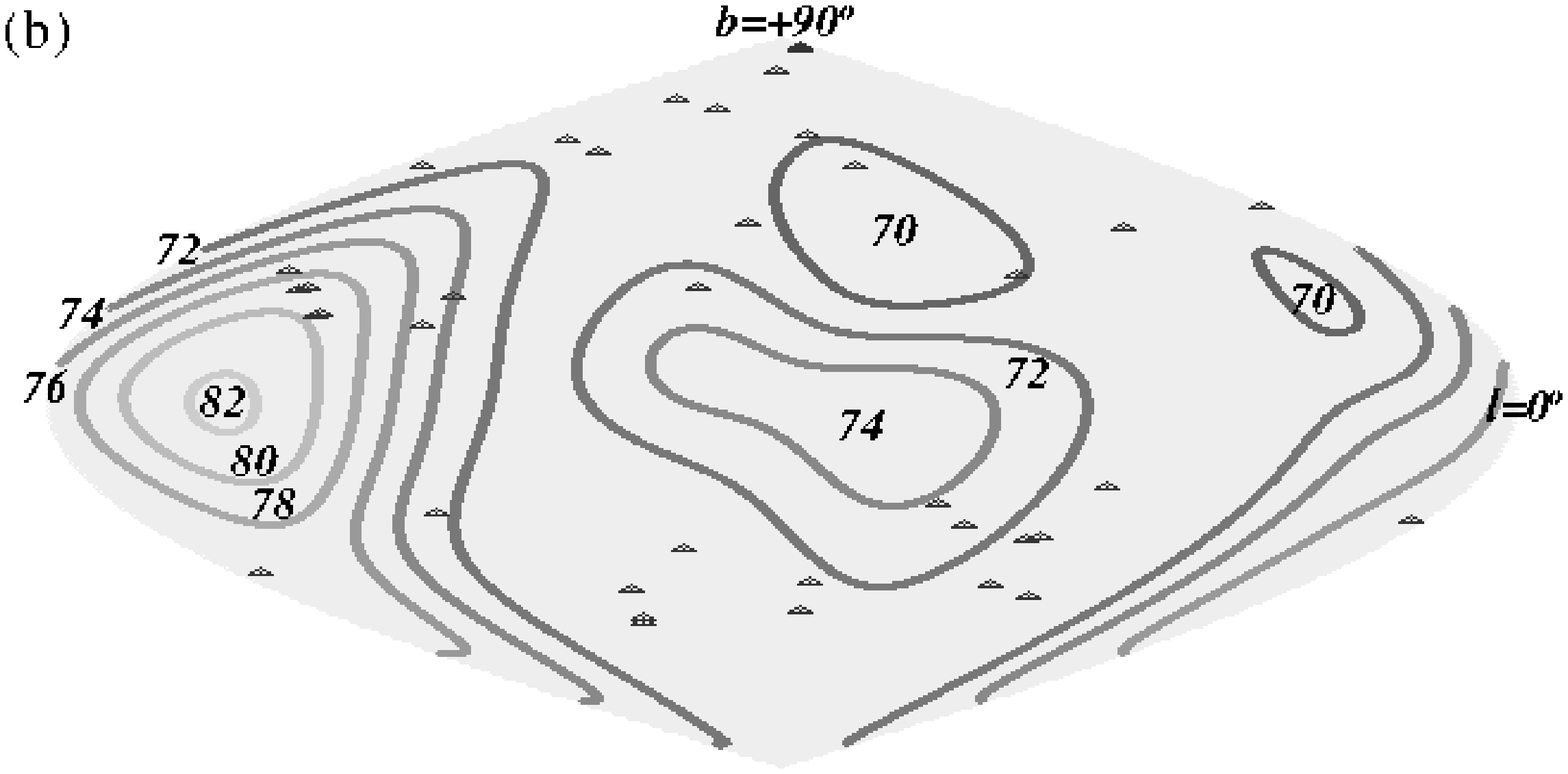}
\caption{Hubble constant contour maps (in Galactic co-ordinates for a sinusoidal projection of the sky) for (a) the 76 \textit{HST} Key Project $H_0$ values and (b) 57
comparison $H_0$ values.
Positions of the actual data points are indicated by triangles, and the contours range from low (dark) to high (light) values of $H_0$ (in km s$^{-1}$
Mpc$^{-1}$) as indicated.}
\end{figure}
                                                                                                                                                                                                                  
In Fig.\ 1 contour maps appear (in Galactic co-ordinates) for the
\textit{HST} Key Project data and the comparison data. The extrema are in similar
directions for the two data sets and are similar in magnitude, except that one of
the maxima in the comparison map is weaker. While the comparison map may not serve
as a completely definitive cross-check of the reality of the variations observed in
the \textit{HST} Key Project map, it certainly seems to show some agreement, and it
does not differ from the \textit{HST} Key Project map as much as would be expected
if the variation in $H_0$ were due to uncertainties in the grid $H_0$ values alone.
This suggests the variation in the maps exists independently of the uncertainties in
the particular determinations of $H_0$. Since there is more uncertainty in the
comparison map than the Key Project map, the differences between the maps probably
mostly reflect errors in the comparison map rather than the Key Project map. 

If the Pavo, Pavo 2, and Ursa Major determinations are removed from the \textit{HST}
Key Project data (due to their discrepant values), the map looks largely unchanged 
from the original:
the primary maximum goes down slightly to 80 km s$^{-1}$ Mpc$^{-1}$ and the 
primary minimum goes up slightly to 71 km s$^{-1}$ Mpc$^{-1}$ while the secondary
extrema remain the same, so that the secondary extrema now become the primary 
extrema in this case.
If PGC 39724 and NGC 4709 are removed from the comparison data, again the map is 
largely unchanged: the primary maximum is slightly lower at just under 79 km 
s$^{-1}$ Mpc$^{-1}$, while the other extrema appear unchanged.
Looking at Fig.\ 1, it is apparent that the extrema tend not to be centred on the
lines-of-sight to the individual data determinations, so the extrema are resulting
from trends in the data, rather than from specific high or low $H_0$ values.

\subsection{Magnitude of variation}
                                                                                                                                                                                                                  
As was previously discussed in Sect.\ 3.1, the Gaussian smearing lessens the range
of any real variation. Using Gaussians with successively smaller standard deviations
of $20^\circ$, $15^\circ$, and $10^\circ$, as the standard deviation gets smaller,
the extrema are picked out with less smearing, but errors in the data also start to
have a greater impact. At $10^\circ$, the range of variation is greater than 30 km
s$^{-1}$ Mpc$^{-1}$, but the grid values are mostly being determined by individual
$H_0$ values so the errors in the grid values approach those of the $H_0$
determinations. The actual extrema are separated by 38 km s$^{-1}$ Mpc$^{-1}$, which
may be an underestimate to the magnitude of any real variation if the standard
deviation could approach zero, but the errors have already become so large that they
are likely making the range of variation appear artificially large as it is. Being
conservative, the range of variation appears to be $\sim$30 km s$^{-1}$ Mpc$^{-1}$.
Thus, large and small values for the standard deviation yield complementary aspects
of the map: smaller standard deviations smooth out the range of variation less,
while larger standard deviations yield the overall trend in the map with less error. 
                                                                                                                                                                                                                  
One may question whether the variation is statistically significant or whether it is
likely that this much variation would be found in the map due to measurement errors
in $H_0$ alone. Assuming that the 1--$\sigma$ random errors are accurate, the
significance of the result can be tested statistically by using Monte Carlo
simulations of the data. This is accomplished using two different methods: one which
assumes the systematic differences between the mean $H_0$ values are real for each
of the 4 methods used to derive $H_0$, and one which assumes that the systematic
differences stem from variations in $H_0$ or errors. The first method involves
calculating a new value of $H_0$ for each position by using the weighted mean value
of $H_0$ corresponding to the actual distance method that was used to obtain the
real $H_0$ value, and then adding Gaussian deviates (varied over a 3--$\sigma$
range) to the data using the weighted mean 1--$\sigma$ error for the corresponding
method to simulate the expected scatter. The second method involves using the
weighted mean value of $H_0$ for all the data for the $H_0$ value at each position
and adding Gaussian deviates to the values according to the actual 1--$\sigma$
errors for each data point. 

For 10,000 different sets of simulated data for each Monte Carlo method, $H_0$ maps
are calculated and the differences between the maximum and minimum $H_0$ grid point
values are found. Comparing the random sky variations with the observed variations
depends strongly on the value of the standard deviation used to smear the values
over the sky. This is because smaller standard deviations smooth out the errors less
and make it more likely to find greater variation in the maps. For the first method,
the results are that the magnitude of variation is only as great as in the real data
5.61\% of the time for a standard deviation of $45^\circ$ (which yields a range of
6.3 km s$^{-1}$ Mpc$^{-1}$), 9.90\% of the time for a standard deviation of
$35^\circ$ (which yields a range of 8.8 km s$^{-1}$ Mpc$^{-1}$), or 37.47\% of the
time for a standard deviation of $25^\circ$ (which yields a range of 12.9 km
s$^{-1}$ Mpc$^{-1}$). The results for the second method are respectively 3.82\%,
4.93\%, or 13.08\%. Thus, assuming the realistic case is somewhere between the two
methods, and assuming anything with a less than 5\% chance is ``statistically
significant," it appears that a statistically significant difference of 6 to 9 km
s$^{-1}$ Mpc$^{-1}$ can be demonstrated with the broader values of the standard
deviation. 
                                                                                                                                                                                                                  
While the above Monte Carlo methods should give a reasonable measure of the
statistical significance of the variation, they depend on the 1--$\sigma$
uncertainties reported for the data being accurate. Independent of the error bars,
one can study whether the values are correlated in position on the sky, with higher
values tending to be near higher values and lower values tending to be near lower
values. Randomly reassigning the $H_0$ values to the actual $H_0$ positions on the
sky and computing maps for several randomizations reveals whether the actual map has
more variation than just the scatter in the data should produce. If there is real
variation in the data, this method will not yield accurate measures of the
statistical significance, since there will be extra scatter in the data that will
tend to allow more variation in the map than errors alone should produce. However,
this method at least yields an upper limit to the likelihood that as much variation
could occur in the map if there were no real variation in the data. 
                                                                                                                                                                                                                  
For 10,000 randomizations of the above method, the results are that the range of
variation is only as great as for the real map 14.78\% of the time for a standard
deviation of $45^\circ$, 21.55\% of the time for a standard deviation of $35^\circ$,
or 42.10\% of the time for a standard deviation of $25^\circ$. Thus, even if one is
a complete skeptic that there is no real variation in the data, there appears to be
more variation than there should be, although of weaker significance than the above
Monte Carlo methods found. On the other hand, if there is real variation in the
data, these are upper bounds on the percentages, so this weaker significance is not
inconsistent with the results of the above methods. Since this method does not
depend on uncertainties, it can also be applied to the comparison data. For the
complete set of 133 data points, the results are that the variation is only as great
1.16\% of the time for a standard deviation of $45^\circ$ (which yields a range of
6.6 km s$^{-1}$ Mpc$^{-1}$), 2.60\% of the time for a standard deviation of
$35^\circ$ (which yields a range of 8.9 km s$^{-1}$ Mpc$^{-1}$), or 19.29\% of the
time for a standard deviation of $25^\circ$ (which yields a range of 12.6 km
s$^{-1}$ Mpc$^{-1}$). Thus, the difficulty in producing coincidental correlations in
larger sets of data allows the conglomerate data set to substantiate a statistically
significant variation of $\sim$9 km s$^{-1}$ Mpc$^{-1}$. 
                                                                                                                                                                                                                  
\subsection{Directional uncertainty and distance dependence}
                                                                                                                                                                                                                  
\begin{figure}
\includegraphics[width=\hsize]{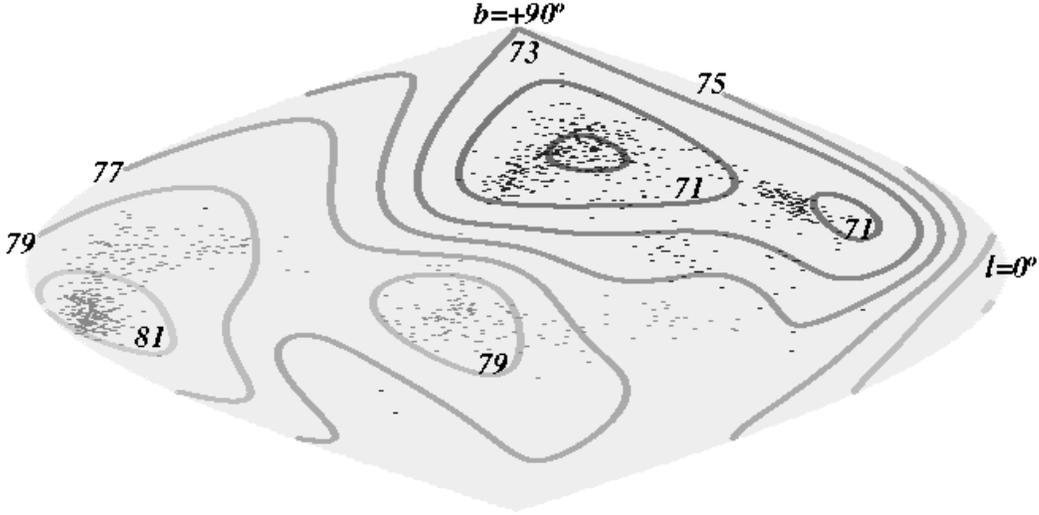}
\caption{Hubble constant contour map of Fig.\ 1(a) with 500 random extrema for maps calculated with Gaussian deviates for the 76 \textit{HST} Key Project
$H_0$ values.
Positions of the minima and maxima are indicated by dark and light dots respectively, and the contours range from low (dark) to high (light) values of $H_0$
(in km s$^{-1}$ Mpc$^{-1}$) as indicated.}
\end{figure}
                                                                                                                                                                                                                  
The directional uncertainty of the extrema in the map can be tested by adding
Gaussian deviates (in a 3--$\sigma$ range) to the 76 data points, computing the
$H_0$ grid map, and seeing how much this affects the positions of the maximum and
minimum $H_0$ grid values in the map. The extrema for each of 500 randomizations are
plotted in Fig.\ 2. Each randomization is calculated with grid separations of
$0.5^\circ$ so that the extrema can be plotted with less granularity. It can be seen
that the probability distribution for the positions of the extrema is not the same
in all directions, but primary and secondary extrema exist with directional
uncertainties of order $10^\circ$ to $20^\circ$. 
                                                                                                                                                                                                                  
To test the influence of depth on the map, an additional weighting factor for
distance is added in the computation of the grid maps. Maps are computed for various
nominal distances by weighting the data according to the fraction of the distance
that is shared in common with a given nominal distance. For distances more distant
than the nominal distance, the weightings go as the ratio of nominal distance to
object distance. For distances closer than the nominal distance, the weightings go
as the ratio of object distance to nominal distance. 

\begin{figure*}
\begin{minipage}[b]{85mm}
\includegraphics[width=\hsize]{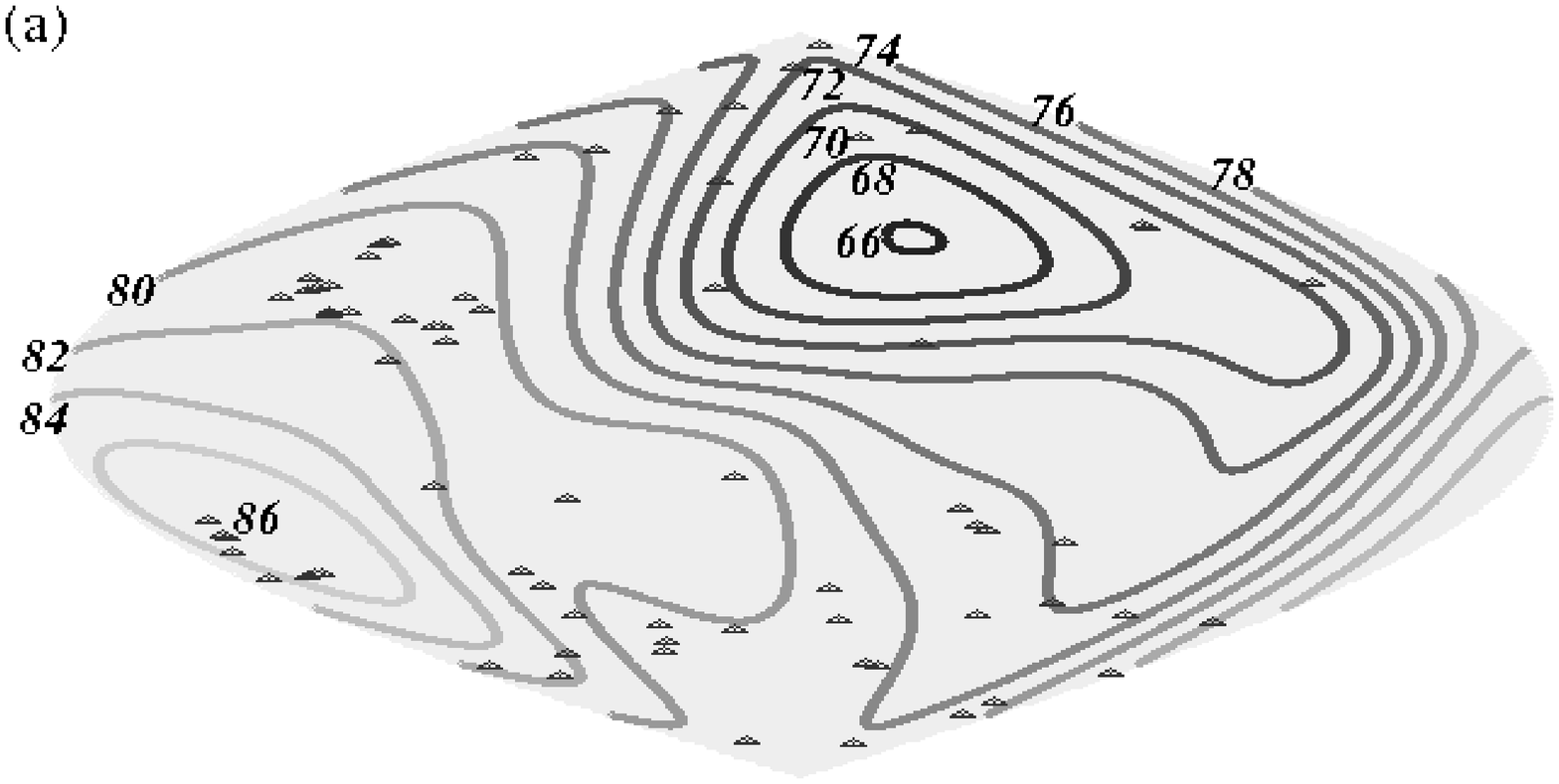}
\end{minipage}
\hfill
\begin{minipage}[b]{85mm}
\includegraphics[width=\hsize]{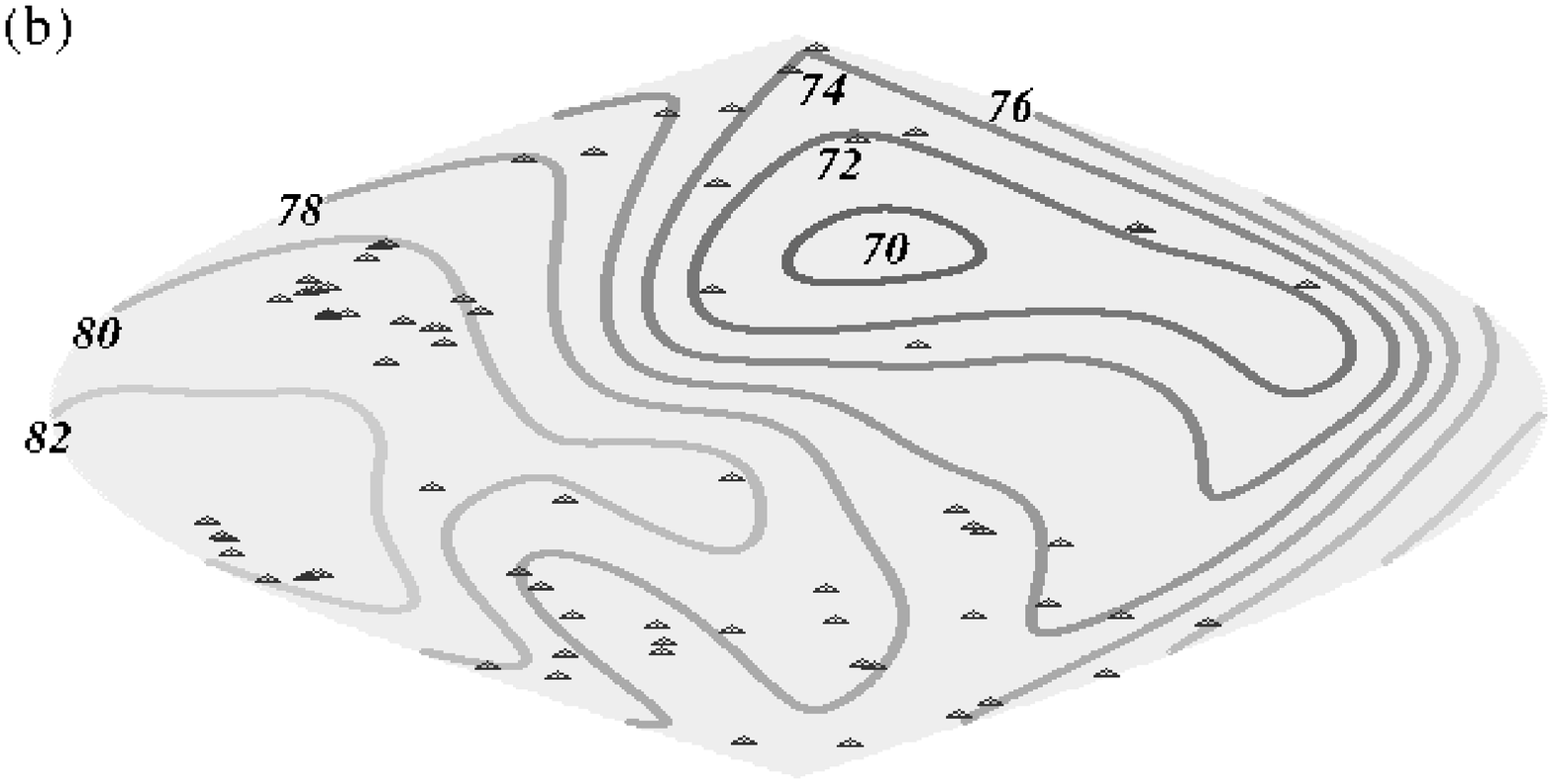}
\end{minipage}
\begin{minipage}[b]{85mm}
\includegraphics[width=\hsize]{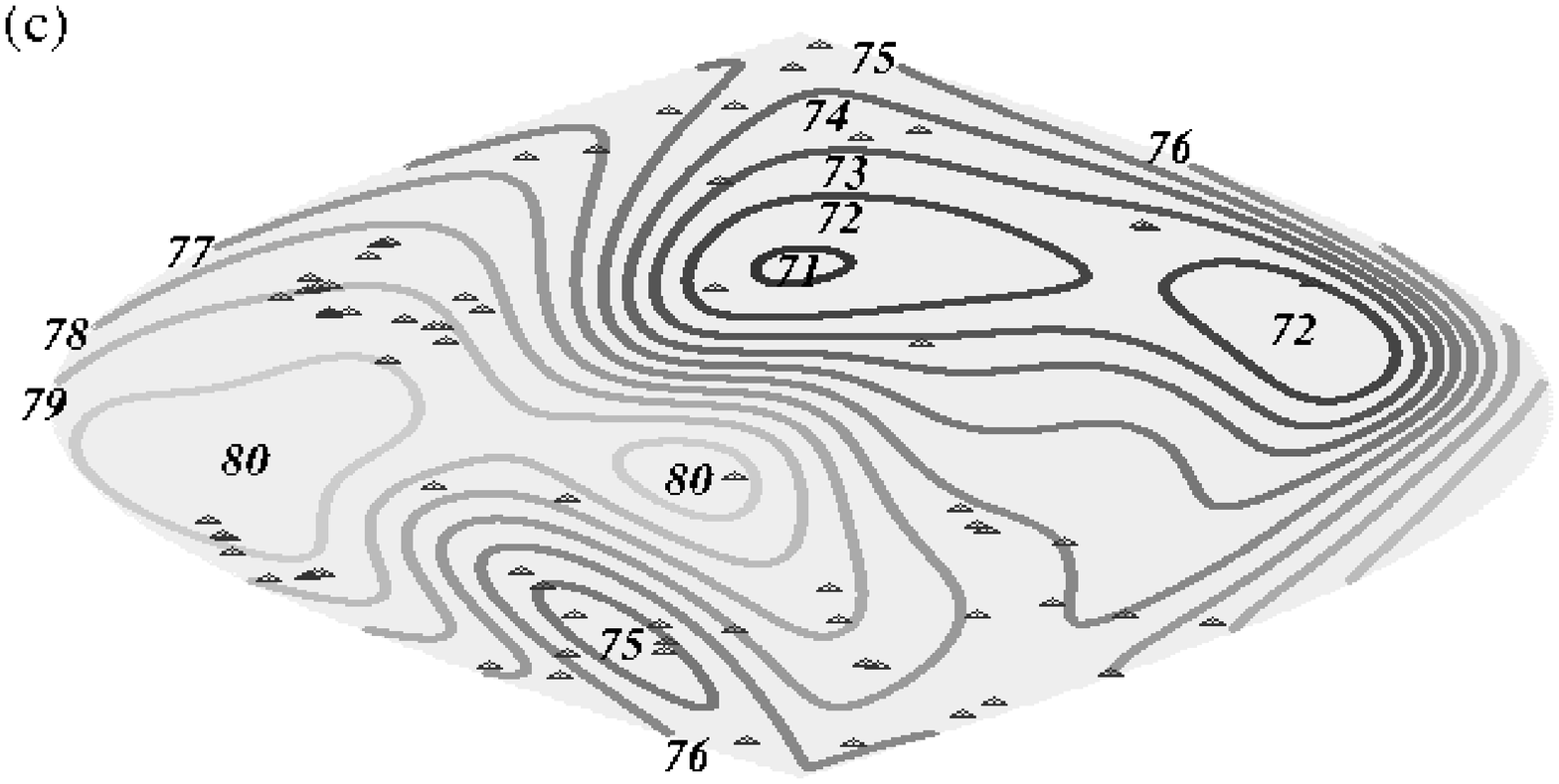}
\end{minipage}
\hfill
\begin{minipage}[b]{85mm}
\includegraphics[width=\hsize]{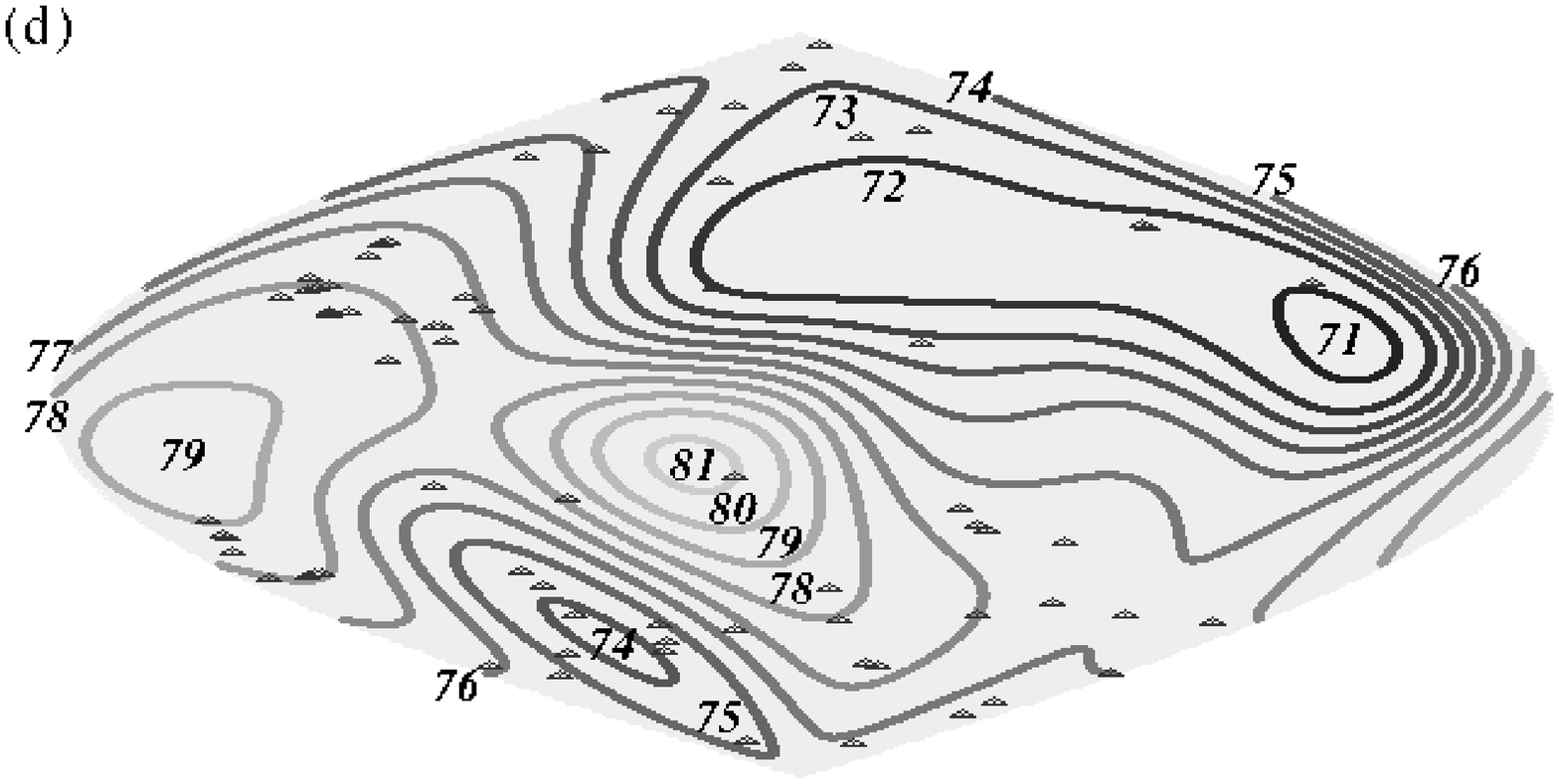}
\end{minipage}
\begin{minipage}[b]{85mm}
\includegraphics[width=\hsize]{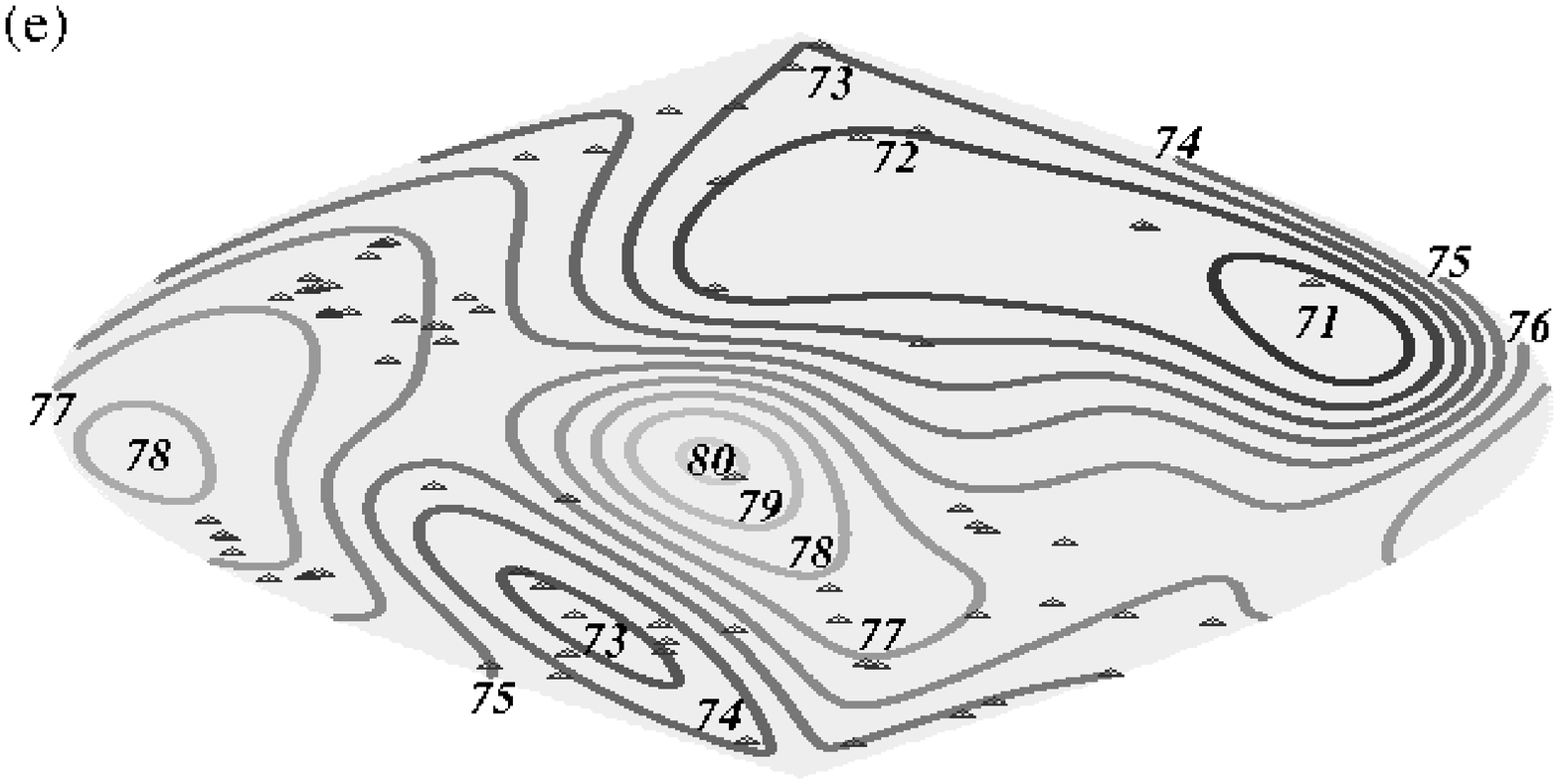}
\end{minipage}
\hfill
\begin{minipage}[b]{85mm}
\includegraphics[width=\hsize]{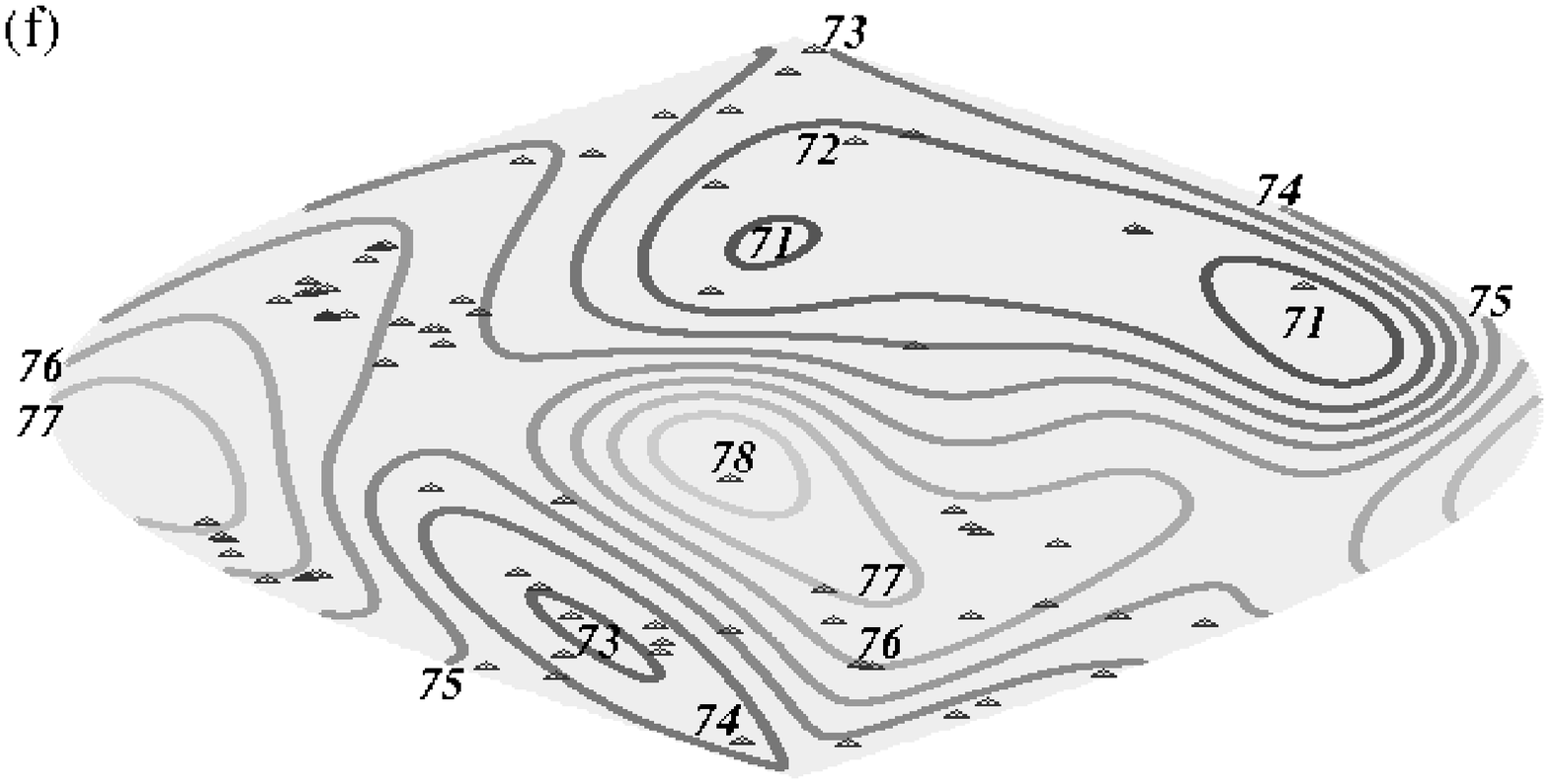}
\end{minipage}
\caption{Hubble constant contour maps (in Galactic co-ordinates for a sinusoidal projection of the sky) for the 76 \textit{HST} Key Project $H_0$ values weighted for nominal distances of (a) 30 Mpc, (b) 
50 Mpc, (c) 80 Mpc, (d) 120 Mpc, (e) 180 Mpc, and (f) 300 Mpc.
Positions of the actual data points are indicated by triangles, and the contours range from low (dark) to high (light) values of $H_0$ (in km s$^{-1}$ Mpc$^{-1}$) as indicated.}
\end{figure*}
                                                                                                                                                                                                                  
About half of the \textit{HST} Key Project data are for objects between 13 and 70
Mpc, with the other half between 70 and 467 Mpc. In Fig.\ 3 distance-weighted maps
appear for 6 nominal distances: 30 Mpc, 50 Mpc, 80 Mpc, 120 Mpc, 180 Mpc, and 300
Mpc. It is apparent that nearby, one pair of extrema from Fig.\ 1 dominates, and the
magnitude of variation of this pair falls off with distance. The minimum is near
($\alpha = 9^h30^m$, $\delta = +70^\circ$), and the maximum is near ($\alpha =
19^h30^m$, $\delta = -70^\circ$). Less apparent is that the grid values of $H_0$ at
the secondary extrema from Fig.\ 1 remain roughly constant with distance and
dominate only at the greatest distances where the range of variation from the first
pair of extrema has become small enough. The secondary minimum is near ($\alpha =
18^h0^m$, $\delta = +15^\circ$), and the secondary maximum is near ($\alpha =
5^h30^m$, $\delta = +5^\circ$). 
                                                                                                                                                                                                                  
It should be noted that more distant values of $H_0$ will always sample local
expansion as well, and while the converse is not true, the weighting method allows
values of $H_0$ for the full range of distances to affect all the maps to some
degree. The extrema that dominate locally appear to imply some sort of local effect,
which could be interpreted as a dipole due to a bulk flow of a local sample volume
with respect to the CMB frame, since the extrema are roughly opposite on the sky.
The secondary extrema remain constant with distance, and are also roughly opposite
of each other on the sky, perhaps suggesting an independent bulk flow of a 
larger-scale volume. 

Removing all the data points from the \textit{HST} Key Project data set that have
an uncertainty greater than 8 km s$^{-1}$ Mpc$^{-1}$ and producing a map with
the remaining 44 points yields a map quite similar to the most distant maps of
Fig.\ 3.
The remaining data essentially consist of type Ia supernova measurements and a few
other surface brightness fluctuation and fundamental plane measurements that are 
for distances about 50 Mpc and greater.
The map ranges from 68 km s$^{-1}$ Mpc$^{-1}$ at the minimum to 81 s$^{-1}$ Mpc$^{-1}$
at the maximum, and the primary extrema of Fig.\ 1 now appear even weaker than
in Fig.\ 3(f).
Since these data mostly have uncertainties of less than 3 km s$^{-1}$ Mpc$^{-1}$,
this provides stronger support for the reality of the extrema observed on large scales
in Fig. 3.

It is difficult to make an all-sky map from only 76 points: weighting for distance
only makes it more difficult to make meaningful all-sky maps. Thus, to test the
significance of the observed variation with distance,
the \textit{HST}
Key Project data are combined with the comparison data, and then divided into two
sets according to distance. 
Randomly reassigning the $H_0$ values to the data positions for the nearby group of
67 data points (to get a limit on the statistical significance) yields the result
that as much variation occurs only 0.16\% of the time for a standard deviation of
$35^\circ$ (which yields a range of variation of 18.8 km s$^{-1}$ Mpc$^{-1}$) or
7.20\% of the time for a standard deviation of $25^\circ$ (which yields a range of
variation of 25.3 km s$^{-1}$ Mpc$^{-1}$). Likewise, the more distant group of 66
data points gives respectively 24.39\% (corresponding to a range of variation of 6.0
km s$^{-1}$ Mpc$^{-1}$) or 23.05\% (corresponding to a range of 11.1 km s$^{-1}$
Mpc$^{-1}$). Thus, given that these percentages are upper limits, while the more
nearby variation demonstrates that a 19 km s$^{-1}$ Mpc$^{-1}$ is certainly
significant, the more distant variation can not be confirmed to be significant.

Unfortunately, there does not appear to be a reasonable way to use the
1--$\sigma$ errors from the combined data sets to properly determine the
statistical significance as in Sect.\ 3.2, as the errors reported for the
comparison data are reported inconsistently and do not share the same
systematic errors. However, just using the 36 type Ia supernova values,
since they have small random errors and will share systematic errors, and
since they tend to be the most distant values, yields interesting results.
It is found that as much variation occurs only 0.00\% of the time for a
standard deviation of $35^\circ$ (which yields a range of variation of 6.7
km s$^{-1}$ Mpc$^{-1}$), 0.02\% of the time for a standard deviation of
$25^\circ$ (which yields a range of variation of 9.7 km s$^{-1}$
Mpc$^{-1}$), and 0.12\% of the time for a standard deviation of $15^\circ$
(which yields a range of variation of 12.9 km s$^{-1}$ Mpc$^{-1}$). If the
1--$\sigma$ errors are truly accurate for the type Ia supernova
measurements, it means it is essentially impossible to achieve this much
variation by chance. To the extent that these 1--$\sigma$ errors can be
trusted, it suggests the variation seen on large scales is statistically
significant. 
                                                                                                                                                                                      
\section{Discussion}
                                                                                                                                                                                                                  
\begin{figure*}
\begin{minipage}[b]{85mm}
\includegraphics[width=\hsize]{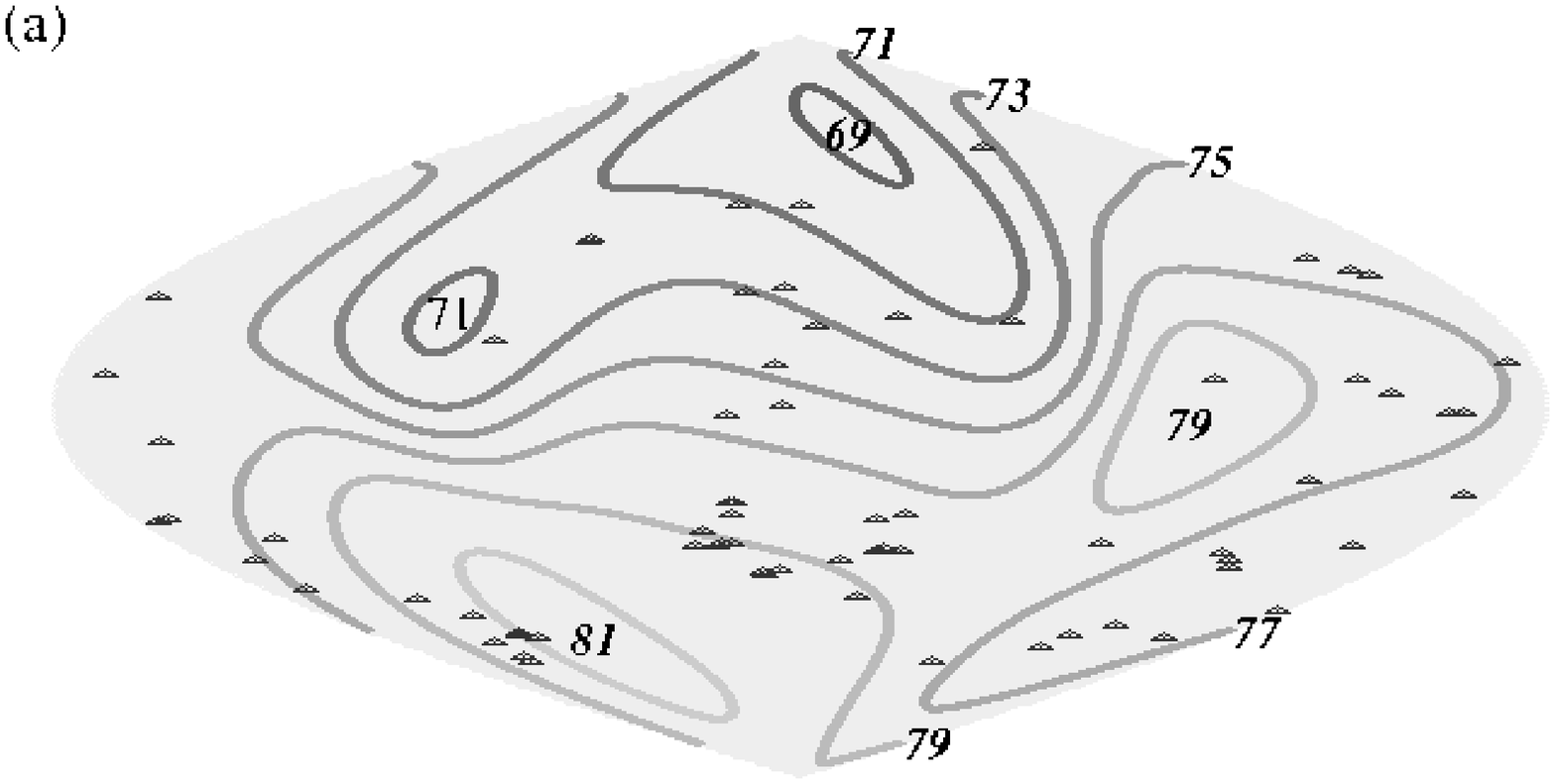}
\end{minipage}
\hfill
\begin{minipage}[b]{85mm}
\includegraphics[width=\hsize]{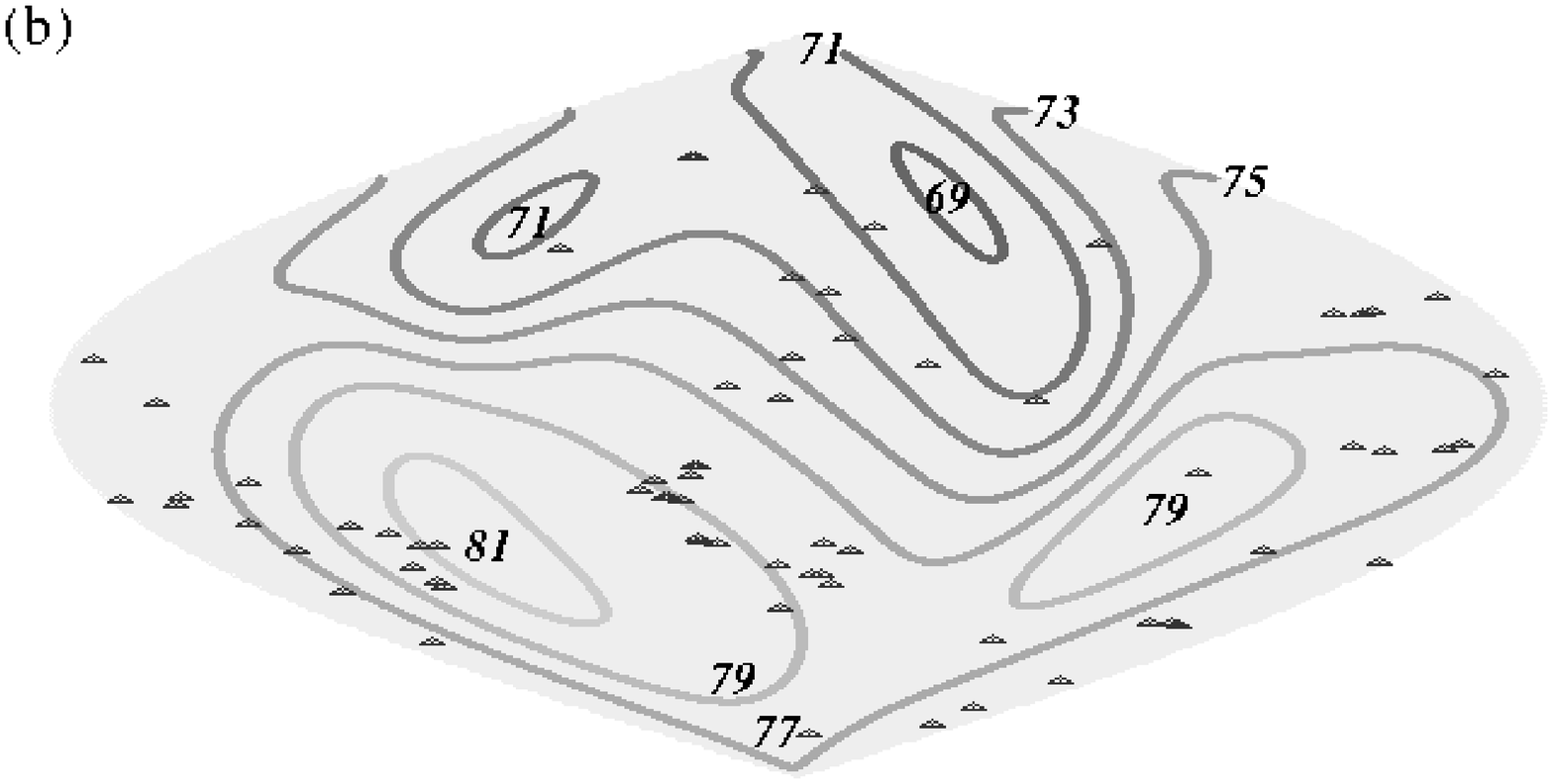}
\end{minipage}
\begin{minipage}[b]{85mm}
\includegraphics[width=\hsize]{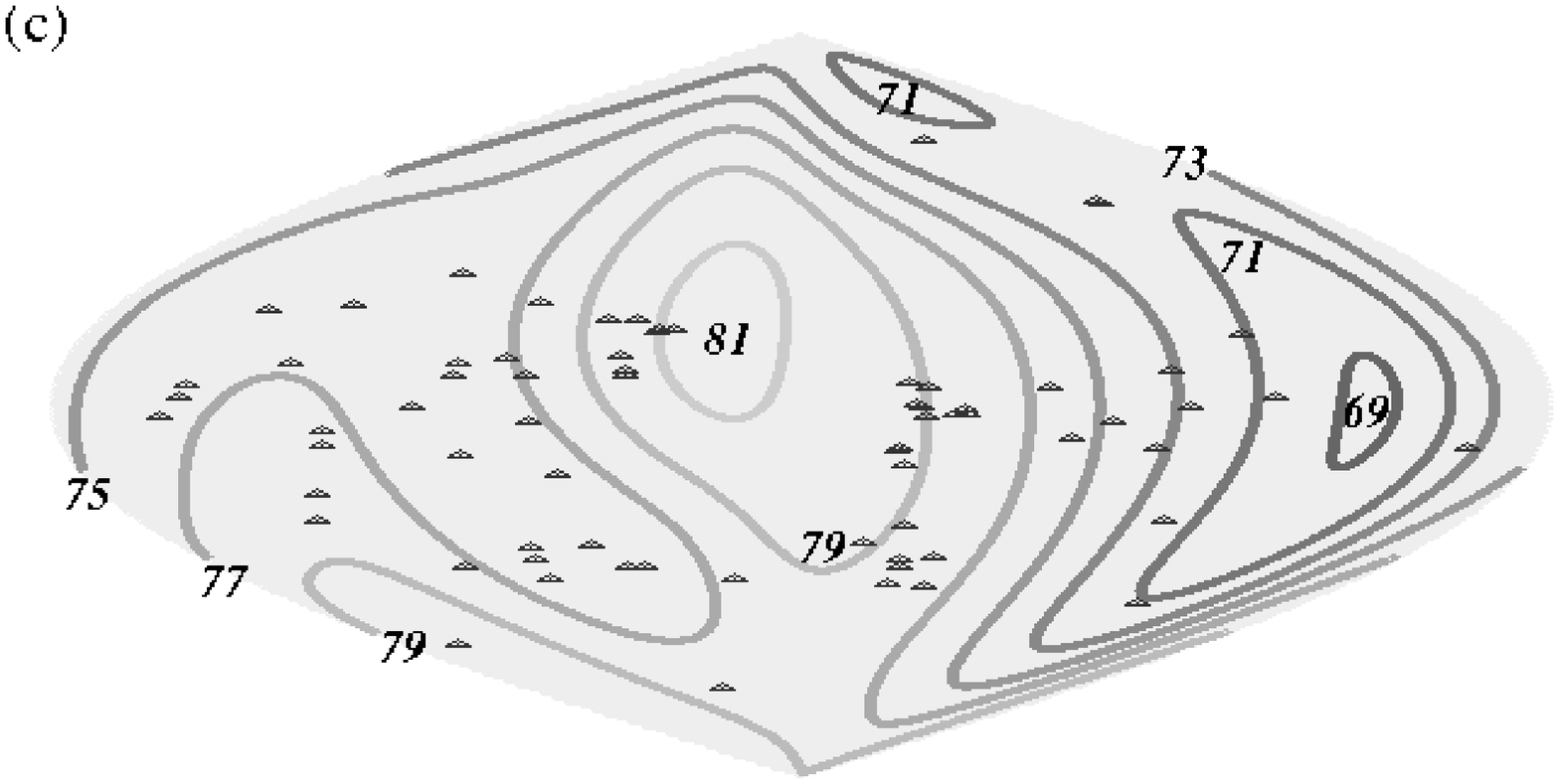}
\end{minipage}
\hfill
\begin{minipage}[b]{85mm}
\includegraphics[width=\hsize]{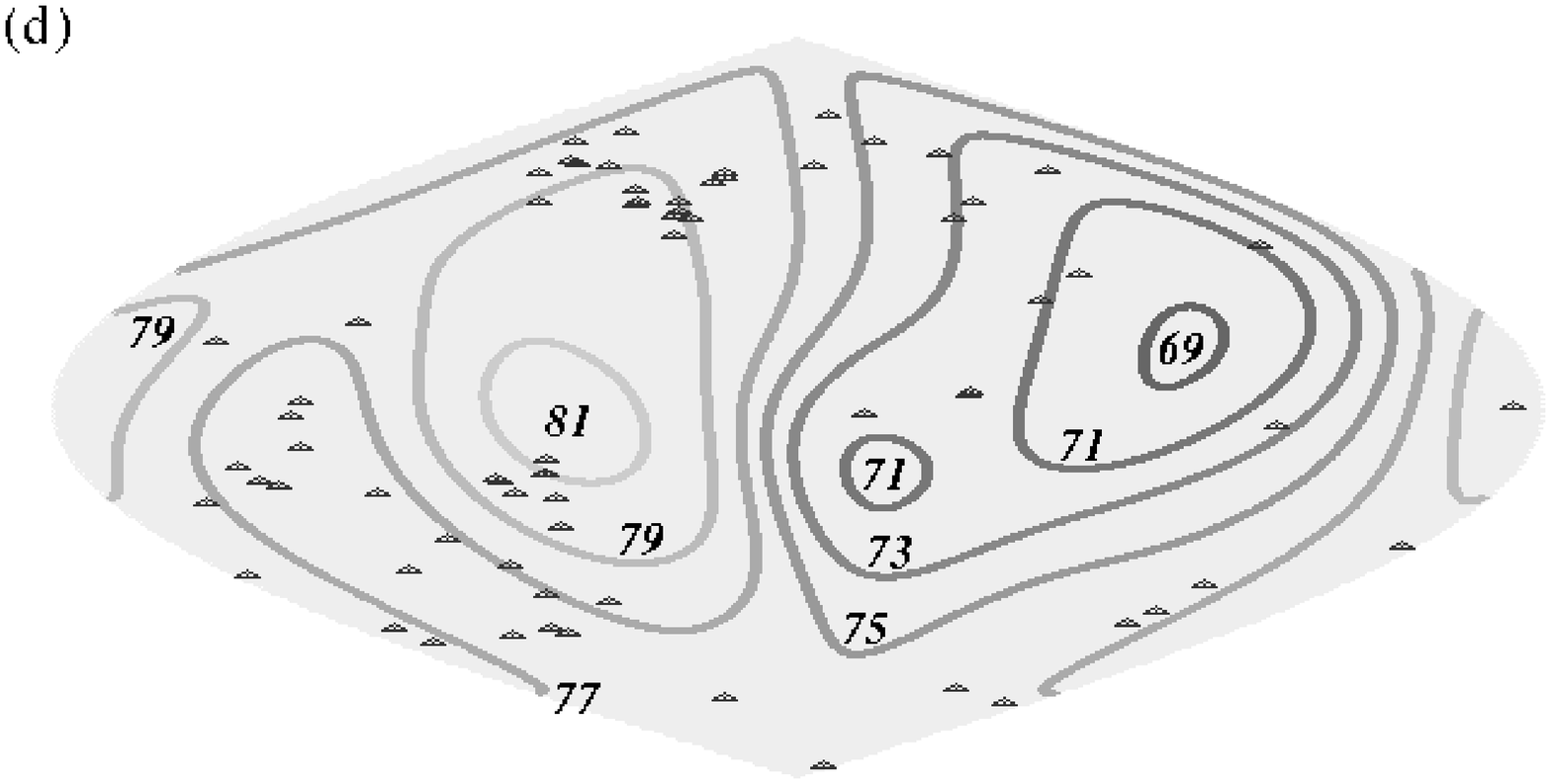}
\end{minipage}
\caption{Hubble constant contour map of Fig.\ 1(a) transformed to (a) celestial co-ordinates, (b) ecliptic co-ordinates, (c) supergalactic co-ordinates,
and (d) the CMB frame.  The poles of the CMB frame are defined by the dipole in the CMB in the heliocentric frame, and the longitude right to
left is defined such that the North Celestial Pole is at $90^\circ$ longitude.
Positions of the actual data points are indicated by triangles, and the contours range from low (dark) to high (light) values of $H_0$ (in km s$^{-1}$
Mpc$^{-1}$) as indicated.}
\end{figure*}
                                                                                                                                                                                                                
The variation does not appear to be an artifact of Galactic dust, since there is no
consistent difference looking in or out of the plane of the Galaxy. In fact, the
overall structure in the map is inconsistent with the distribution of dust in the
COBE dust maps \citep[Schlegel et al.][]{Sch98}, so it seems unlikely that the
observed variation in $H_0$ could be due to poor corrections for dust in our own
Galaxy. In Fig.\ 4 the \textit{HST} Key Project map of Fig.\ 1(a) has been
transformed respectively to celestial co-ordinates, ecliptic co-ordinates,
supergalactic co-ordinates, and a CMB-dipole-oriented frame of reference for
comparison. Figures 4(a) and 4(b) demonstrate that the extrema do not appear to be
an artifact of any local frame of reference. 
                                                                                                                                                                                                                  
The supergalactic map is interesting because the extrema that predominate locally
are near the supergalactic plane, suggesting they may be associated with local
conglomerations of matter (which tend to be arranged along the supergalactic plane).
Meanwhile, the extrema that dominate farther out are oriented closer to the
supergalactic poles, so if these extrema are a real phenomenon, we may be getting a
clear view looking out of the supergalactic plane and observing effects associated
with a much larger scale. 
                                                                                                                                                                                                                  
The CMB-frame map is oriented with its north pole in the direction of the Sun's
peculiar velocity with respect to the CMB\@. From the map it is apparent that there
is little difference between directions oriented along the CMB dipole. All of the
extrema are near the equator of the map, roughly perpendicular to the CMB dipole.
Since the $H_0$ values have been corrected to be in the CMB frame of reference, this
suggests there is no error due to the correction for the CMB dipole: it seems too
good in fact. Interestingly, this variation seems to be in agreement with the CMB
anisotropy observed by Tegmark et al.\citet{Teg03}, who found that the CMB
quadrupole and octupole are aligned such that the
extrema are in a plane roughly perpendicular to the direction of the dipole. If our 
dipole motion with respect to the CMB is related to the existence of the $H_0$/CMB
anisotropy perpendicular to this direction, it suggests this motion and the
variations related to it may both be stemming from large-scale inhomogeneities. 
Inoue and Silk\citet{Ino06} have suggested the CMB quadrupole and octupole alignment 
can be explained by a pair of voids a few hundred Mpc distant in the 
direction $(l=330^\circ,b=-30^\circ)$, which is interestingly in the direction of 
the $H_0$ maxima in our maps.

\begin{figure}
\includegraphics[width=\hsize]{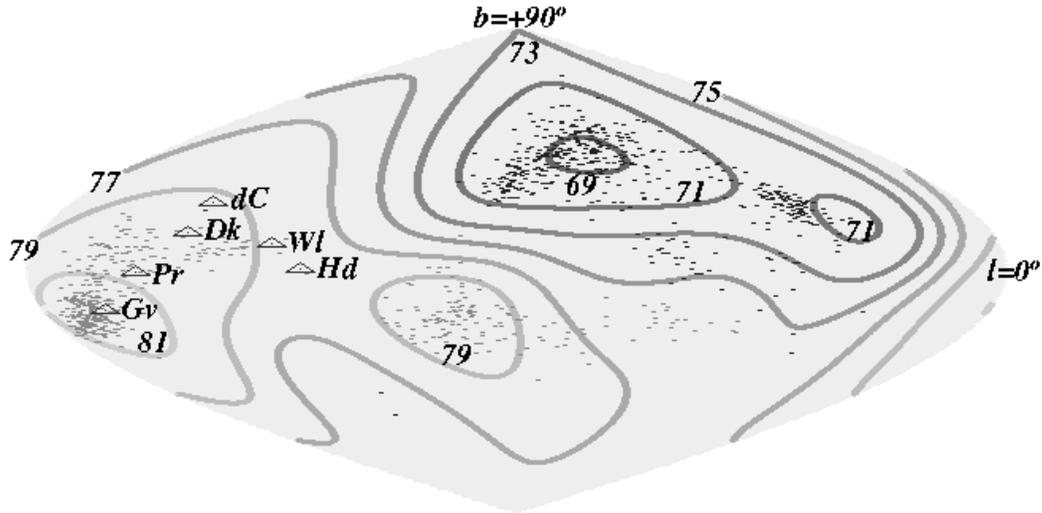}
\caption{Hubble constant contour map of Fig.\ 2 with recent determinations of bulk flows (triangles).  From high to low latitude, the bulk flows
are those of da Costa et al.\citet{daC00}, Dekel et al.\citet{Dek99}, 
Willick\citet{Wil99}, Hudson et al.\citet{Hud04}, Parnovsky et al.\citet{Par01}, 
and Giovanelli et al.\citet{Gio98}.
Positions of the minima and maxima are indicated by dark and light dots respectively, and the contours range from low (dark) to high (light) values of $H_0$
(in km s$^{-1}$ Mpc$^{-1}$) as indicated.}
\end{figure}

In Fig.\ 5 recent bulk flow directions are plotted on top of the map of Fig.\ 2. The
bulk flow directions tend to lie on the outskirts of the uncertainty in the maxima,
so they are not totally consistent with the map. They should be, as bulk flows will
be oriented toward directions associated with higher recessional velocities in the
CMB frame and hence the directions of higher $H_0$. 
However, the bulk flow directions do lie
in the higher $H_0$ regions of the map, so they are not that disconcordant. One
thing to note is that the bulk flow directions appear to be sandwiched between the
primary and secondary maxima. This suggests that by only looking at the net flow,
bulk flow studies may be missing the distinction between two separate effects and
missing the actual directions of interest.

\section{Conclusion}
                                                                                                                                                                                                                  
It appears that a statistically significant variation in $H_0$ of at least 9 km
s$^{-1}$ Mpc$^{-1}$ exists in the \textit{HST} Key Project data. 
The approximate directional uncertainty is $10^\circ$ to $20^\circ$. 
Maps weighted for distance appear to indicate two sets of extrema that dominate 
on different distance scales.
Within our supercluster, differences as great as $\sim$35 km s$^{-1}$ Mpc$^{-1}$ are observed,
and these tend to occur near the supergalactic plane with a minimum near ($\alpha =
9^h30^m$, $\delta = +70^\circ$) and a maximum near ($\alpha = 19^h30^m$, $\delta =
-70^\circ$). Beyond our supercluster, differences as great as $\sim$20 km s$^{-1}$
Mpc$^{-1}$ are observed, and these tend to occur away from the supergalactic plane
with a minimum near ($\alpha = 18^h0^m$, $\delta = +15^\circ$) and a maximum near
($\alpha = 5^h30^m$, $\delta = +5^\circ$). 
Within 70 Mpc, a combination of the \textit{HST} Key Project data and the comparison
data shows a statistically significant difference of 19 km s$^{-1}$ Mpc$^{-1}$.
Beyond 50 Mpc, the \textit{HST} Key Project type Ia supernova data 
yield a statistically significant difference of 13 km s$^{-1}$ Mpc$^{-1}$ 
(assuming the reported 1-$\sigma$ errors are reliable).
                                                                                                                                                                                    
Further study of the resilience of this result requires more data specifically
selected for achieving optimal sky coverage. It would also be interesting to have
data for a greater range of distances to see how far out a statistically significant
variation can be detected and over how large a scale the Universe's expansion needs
to be sampled before it appears to become uniform. 

Real variation in $H_0$ is not really unexpected given the degree of structure and
mass inhomogeneity present in the Universe. One implication of this variation is it
also partially explains why $H_0$ has historically been plagued by so much
uncertainty.

We would like to acknowledge useful suggestions and comments from Roberto Abraham, 
Byron Desnoyers Winmill, John Dubinski, Shoko Sakai, John Tonry, Howard Yee, 
and the referee. 
This work was supported by NSERC through a Post-Graduate Scholarship to MLM and 
a Discovery grant to CCD.


\begin{thebibliography}{}

\bibitem[(1997)]{Bau97} Baum, W.~A., Hammergren, M., Thomsen, B., et al., 1997. AJ, 
113, 1483. 
\bibitem[(2003)]{Ben03} Bene, G., Czinner, V., Vas\'{u}th, M., 2003. Preprint 
astro-ph/0308161 v4.
\bibitem[(1991)]{Bil91} Bildhauer, S., Futamase, T., 1991. Gen.\ Relativ.\ Grav., 23, 
1251.
\bibitem[(2000)]{daC00} da Costa, L.~N., Bernardi, M., Alonso, M.~V., et al., 2000. 
ApJ, 537, L81.
\bibitem[(1999)]{Dek99} Dekel, A., Eldar, A., Kolatt, T., et al., 1999. ApJ, 522, 1.
\bibitem[(2000)]{Fer00} Ferrarese, L., Mould, J.~R., Kennicutt, R.~C., Jr., et al., 
2000. ApJ, 529, 745.
\bibitem[(1996)]{For96} Forbes, D.~A., Brodie, J.~P., Huchra, J., 1996. AJ, 112, 2448.
\bibitem[(1996)]{Ford96} Ford, H.~C., Hui, X., Ciardullo, R., Jacoby, G.~H.,
Freeman, K.~C., 1996. ApJ, 458, 455.
\bibitem[(2001)]{Fre01} Freedman, W.~L., Madore, B.~F., Gibson, B.~K., et al., 2001. 
ApJ, 553, 47.   
\bibitem[(1998)]{Gio98} Giovanelli, R., Haynes, M.~P., Freudling, W., da Costa, 
L.~N., 1998. ApJ, 505, L91.
\bibitem[(1997)]{Gre97} Gregg, M.~D., 1997. NewA, 1, 363.
\bibitem[(1995)]{Her95} Herbig, T., Lawrence, C.~R., Readhead, A.~C.~S., Gulkis, 
S., 1995. ApJ, 449, L5.
\bibitem[(1997)]{Hjo97} Hjorth, J., Tanvir, N.~R., 1997. AJ, 482, 68.
\bibitem[(2004)]{Hud04} Hudson, M.~J., Smith, R.~J., Lucey, J.~R., Branchini, E.,
2004. MNRAS, 352, 61.
\bibitem[(2006)]{Ino06} Inoue, K.~T., Silk, J., 2006. ApJ, 648, 23.
\bibitem[(1997)]{Jen97} Jensen, J.~B., 1997. Ph.D. Thesis, Univ.\ Hawaii.
\bibitem[(2001)]{Jen01} Jensen J.~B., Tonry, J.~L., Thompson, R.~I., et al., 2001. 
ApJ, 550, 503.
\bibitem[(2005)]{Kolb05} Kolb, E.~W., Matarrese, S., Notari, A., Riotto, A., 2005. 
Phys.\ Rev.\ D, 71, 023524.
\bibitem[(1998)]{Lau98} Lauer, T.~R., Tonry, J.~L., Postman, M., Ajhar, E.~A., 
Holtzman, J.~A., 1998. ApJ, 499, 577.
\bibitem[(2001)]{Liu01} Liu, M.~C., Graham, J.~R., 2001. ApJ, 557, L31.
\bibitem[(1996)]{Mad96} Madore, B.~F., Freedman, W.~L., Kennicutt, R. C., et al., 
1996. Am.\ Astron.\ Soc.\ Meeting, 189, 108.04.
\bibitem[(1995)]{Mof95} Moffat, J.~W., Tatarski, D.~C., 1995. ApJ, 453, 17.
\bibitem[(1997)]{Mye97} Myers, S.~T., Baker, J.~E., Readhead, A.~C.~S., Leitch, 
E.~M., Herbig, T., 1997. ApJ, 485, 1.
\bibitem[(2001)]{Par01} Parnovsky, S.~L., Kudrya, Y.~N., Karachentseva, V.~E., 
Karachentsev, I.~D., 2001. Astron.\ Lett.\, 27, 765.
\bibitem[(1998)]{Pat98} Paturel, G., Lanoix, P., Teerikorpi, P., et al., 1998. A\&A, 
339, 671.
\bibitem[(1999)]{Per99} Perlmutter, S., Aldering, G., Goldhaber, G., et al., 1999. 
ApJ, 517, 565.
\bibitem[(1955)]{Ray55} Raychaudhuri, A., 1955. Phys.\ Rev., 98, 1123.
\bibitem[(1997)]{Russ97} Russ, H., Soffel, M.~H., Kasai, M., B\"{o}rner, G., 1997. 
Phys.\ Rev.\ D, 56, 2044.
\bibitem[(2000)]{Sak00} Sakai, S., Mould, J.~R., Hughes, S.~M.~G., et al., 2000. ApJ, 
529, 698.
\bibitem[(2001)]{Sak01} Sakai, S., 2001. Private communication.
\bibitem[(1998)]{Sal98} Salaris, M., Cassisi, S., 1998. MNRAS, 298, 166.
\bibitem[(1998)]{Sch98} Schlegel, D.~J., Finkbeiner, D.~P., Davis, M., 1998. ApJ, 
500, 525.
\bibitem[(2003)]{Teg03} Tegmark, M., de~Oliveira-Costa, A., Hamilton, A.~J., 2003. 
Phys.\ Rev.\ D, 68, 123523.
\bibitem[(1997)]{Tho97} Thomsen, B., Baum, W.~A., Hammergren, M., Worthey, G., 
1997. ApJ, 483, 37.
\bibitem[(2000)]{Tul00} Tully, R.~B., Pierce, M.~J., 2000. ApJ, 533, 744.
\bibitem[(1995)]{Whi95} Whitmore, B.~C., Sparks, W.~B., Lucas, R.~A., Macchetto, 
F.~D., Biretta, J.~A., 1995. ApJ, 454, L73.
\bibitem[(1999)]{Wil99} Willick, J.~A., 1999. ApJ, 522, 647.
\bibitem[(2002)]{Zar02} Zaroubi, S., 2002. Preprint astro-ph/0206052 v2.
\bibitem[(1996)]{Zas96} Zasov, A.~V., Bizyaev, D.~V., 1996. Astron.\ Lett., 22, 71.
\bibitem[(1998)]{Zeh98} Zehavi, I., Riess, A.~G., Kirshner, R.~P., Dekel, A., 1998. 
ApJ, 503, 483.

 
\end{thebibliography}
\end{document}